\documentclass[12pt]{JHEP3}

\usepackage{amsmath,amssymb,amsfonts,amsthm}
\usepackage{bm}
\usepackage{graphicx}

\newcommand{\be}{\begin{equation}}
\newcommand{\ee}{\end{equation}}
\newcommand{\ba}{\begin{eqnarray}}
\newcommand{\ea}{\end{eqnarray}}
\def\bs{\begin{subequations}}
\def\es{\end{subequations}}
\def\a{\alpha}
\def\b{\beta}
\def\g{\gamma}
\def\la{\lambda}
\def\k{\kappa}

\def\ve{\varepsilon}
\def\Om{\Omega}
\def\om{\omega}
\def\G{\Gamma}

\def\s{\sigma}
\def\vr{\varrho}
\def\vp{\varphi}
\def\N{\nabla}
\def\cC{{\cal C}}
\def\cE{{\cal E}}
\def\cH{{\cal H}}
\def\cL{{\cal L}}

\def\cM{{\cal M}}

\def\cK{{\cal K}}
\def\cJ{{\cal J}}

\def\p{\partial}
\def\B{\Box}
\newcommand{\Eq}[1]{(\ref{#1})}

\def\rme{e}
\def\rmd{d}
\def\rmi{i}

\title{Quantum field theory, gravity and cosmology in a fractal universe}

\author{Gianluca Calcagni\\
Max Planck Institute for Gravitational Physics (Albert Einstein Institute)\\
Am M\"uhlenberg 1, D-14476 Golm, Germany\\
E-mail: \email{calcagni@aei.mpg.de}}

\date{January 1, 2010}

\abstract{We propose a model for a power-counting renormalizable field theory living in a fractal spacetime. The action is Lorentz covariant and equipped with a Stieltjes measure. The system flows, even in a classical sense, from an ultraviolet regime where spacetime has Hausdorff dimension 2 to an infrared limit coinciding with a standard $D$-dimensional field theory. We discuss the properties of a scalar field model at classical and quantum level. Classically, the field lives on a fractal which exchanges energy-momentum with the bulk of integer topological dimension $D$. Although an observer experiences dissipation, the total energy-momentum is conserved. The field spectrum is a continuum of massive modes. The gravitational sector and Einstein equations are discussed in detail, also on cosmological backgrounds. We find ultraviolet cosmological solutions and comment on their implications for the early universe.}

\preprint{JHEP03(2010)120 \hspace{8.3cm} arXiv:1001.0571}

\begin{document}


\section{Introduction}

The search for a consistent theory of quantum gravity is one of the main issues in the present agenda of theoretical physics. Beside major efforts such as string theory and loop quantum gravity, other independent lines of investigation have recently attracted some attention. Among these, Ho\v{r}ava--Lifshitz (HL) gravity \cite{Hor2,Hor3} is a proposal for a power-counting renormalizable model \cite{Vi09a,Vi09b} which is not Lorentz invariant. Coordinates scale anisotropically, i.e., $[t]=-z$ and $[x^i]=-1$ in momentum units, where $z\geq 3$ is a critical exponent typically fixed at $z=3$. Because of this, the total action can be engineered so that the effective Newton constant becomes dimensionless in the ultraviolet (UV) and higher-order spatial derivatives improve the short-scale behaviour of particle propagators. Due to the presence of relevant operators, the system is conjectured to flow from the UV fixed point to an infrared (IR) fixed point where, effectively, Lorentz and diffeomorphism invariance is restored at classical level.

Another property of HL gravity stemming from the running of the couplings effective dimension is that the spectral dimension $d_{\rm S}$ \cite{AO,RT,AJW,bH} at short scales is $d_{\rm S}\sim 2$ \cite{Hor3}. This is in intriguing accordance with other proposals for quantum gravity such as causal dynamical triangulations \cite{AJL4}, asymptotically safe gravity \cite{LaR5} and spin-foam models \cite{Mod08} (see also \cite{Ben08}). Systems whose effective dimensionality changes with the scale can show fractal behaviour, even if they are defined on a smooth manifold.\footnote{HL gravity with detailed balance possesses a natural fractal structure also because of the appearance of fractional pseudo-differential operators \cite{Lif1,Lif2}. This version of the theory, however, seems to be unviable.} All the above examples incarnate the popular notion that ``the Universe is fractal'' at quantum scales.

Despite the beautiful physics emerging from the HL picture, inspired by critical and condensed-matter systems, it potentially suffers from at least one major problem. Lorentz invariance, one of the best constrained symmetries of Nature, is surrendered at fundamental level. As argued on general grounds \cite{CPSUV,CPS}, even if deviations from Lorentz invariance are classically negligible, loop corrections to the propagator of fields lead to violations several orders of magnitude larger than the tree-level estimate, unless the bare parameters of the model are fine tuned. This expectation \cite{Lif2} is indeed fulfilled for Lifshitz-type scalar models \cite{IRS}. Although supersymmetry might relax the fine tuning \cite{wor}, the present version of HL gravity is clearly under strong pressure, also for other independent reasons.

Motivated by the virtues and problems of HL gravity, it is the purpose of this paper to formulate an effective quantum field theory with two key features. The first is that power-counting renormalizability is obtained when the fractal behaviour is realized at structural level, i.e., when it is implemented in the very definition of the action rather than as an effective property. In other words, we will require not only the spectral dimension of spacetime, but also its UV Hausdorff dimension \cite{Fal03} (which will coincide with $d_{\rm S}$ in our case) to be $d_{\rm H}\sim2$. Secondly, we wish to maintain Lorentz invariance.

Therefore, this proposal is (a) defined on a fractal (in a sense made precise below), (b) Lorentz invariant, (c) power-counting renormalizable, (d) UV finite with no ghost or other obvious instabilities, and (e) causal. A condensed overview of the model was given in \cite{fra1}.

Some of the ingredients we shall use are similar to those found in other recipes (e.g., scalar-tensor theories or models with fractional operators). Their present mixing, however, will hopefully give fresh insight into some aspects of quantum gravity. For example, a running cosmological constant naturally emerges from geometry as a consequence of a deformation of the Poincar\'e algebra.

The plan of the paper is the following. The main idea is introduced in section \ref{fra}. With particular reference to a scalar field theory, a dimensional analysis of the coupling constants is given in section \ref{renor}. Section \ref{scal} is devoted to a scalar field on a Minkowski fractal: its classical equation of motion and dynamics are presented in section \ref{eoh}, where the Hamiltonian formalism is shown to admit both a dissipative and conservative interpretation. The causal propagator of the free field in configuration space is calculated in section \ref{proc}, while its Fourier--Stieltjes transform in momentum space is discussed in section \ref{prok}. We outline the gravitational sector in section \ref{gravy}. Einstein and cosmological equations are derived in sections \ref{einseqs} and \ref{cosmo}, respectively, where cosmological solutions are found and analyzed. Section \ref{disc} contains concluding remarks and a discussion on open issues and future developments.


\section{Fractal universe}\label{fra}

In HL gravity, one requires that time and space coordinates scale anisotropically. On one hand, this leads to a running scaling dimension of the couplings and an effective two-dimensional phase in the UV. On the other hand, anisotropic scaling gives rise to higher-order spatial operators and a non-Lorentz-invariant action. It turns out that we can achieve the first result (and avoid the second) by maintaining isotropic scaling,
\be\label{iso}
[x^\mu]=-1\,,\qquad \mu=0,1,\dots,D-1\,,
\ee
while replacing the standard measure with a nontrivial Stieltjes measure,
\be
\rmd^D x\to \rmd\vr(x)\,,\qquad [\vr]=-D\a\neq -D\,.
\ee
Here $D$ is the topological (positive integer) dimension of embedding (abstract) spacetime and $\a>0$ is a parameter.\footnote{The Hausdorff dimension of a set is greater than or equal to its topological dimension but the situation one has in mind here is a physical spacetime (the fractal) embedded in an ambient $D$-dimensional manifold $\cM$. All physics takes place in the fractal and there are no observers in the ``bulk'' $\cM$. Given this picture, one can interpret the present model as ``diffusion of spacetime'' in an embedding manifold.} What kind of measure can we choose? A two-dimensional small-scale structure is a desirable feature of renormalizable spacetime models of quantum gravity, and the most na\"ive way to obtain it is to let the effective dimensionality of the universe to change at different scales. A simple realization of this feature is via fractional calculus and the definition of a fractional action.\footnote{Another route, which we shall not follow here, is to define particle physics directly on a fractal set with general Borel probability measure $\vr$. This was done in \cite{Svo87} (and \cite{Ey89a,Ey89b} on Sierpinski carpets) for a quantum field theory on sets with Hausdorff dimension $4-\epsilon$ very close to 4. The model in \cite{Svo87} has many aspects of dimensional regularization \cite{BG72,tV}, one difference being that the parameter $\epsilon$ is taken to be physical.}

To begin with, we quote the following results in classical mechanics. In \cite{tat95}, empirical evidence was given that the Hausdorff dimension of a random process (Brownian motion) described by a fractional differintegral is proportional to the order $\a$ of the differintegral; the same relation holds for deterministic fractals, and in general the fractional differintegration of a curve changes its Hausdorff dimension as $d_{\rm H}\to d_{\rm H}+\a$ (see also \cite{MH}). Moreover, integrals on net fractals can be approximated by the left-sided Riemann--Liouville fractional integral of a function $L(t)$ \cite{RYS,YRZ,QL,RQLW,RLWQ},
\ba
\int_0^{\bar t}\rmd\vr(t)\, L(t) &\propto& {}_0I_{\bar t}^{\a}L(t)\label{net}\\
&\equiv& \frac{1}{\Gamma(\a)}\int_0^{\bar t}\rmd t\,(\bar t-t)^{\a-1} L(t)\,,\label{rli}\\
\vr(t) &=& \frac{{\bar t}^\a-(\bar t-t)^\a}{\Gamma(\a+1)}\,,\label{rli2}
\ea
where $\bar t$ is fixed and the order $\a$ is (related to) the Hausdorff dimension of the set \cite{RYS,Nig92}. The approximation in eq.~\Eq{net} is valid for large Laplace momenta and can be refined to better describe the full structure of the Borel measure $\vr$ characterizing the fractal set. In the latter case, integration on the set is approximated by a sum of fractional integrals \cite{QL}.

Different values of $0<\a\leq 1$ mediate between full-memory ($\a=1$) and Markov processes ($\a=0$), and in fact $\a$ roughly corresponds to the fraction of states preserved at a given time $\bar t$ during the evolution of the system \cite{RYS,RLWQ,Nig92}. Applications of fractional integrals range from statistics, diffusing or dissipative processes with residual memory \cite{RLWQ}, such as weather and stochastic financial models \cite{Mis08}, to system modeling and control in engineering \cite{Das08}.

Noticing that a change of variables $t\to\bar t-t$ transforms eq.~\Eq{rli} into the form
\be
\frac{1}{\Gamma(\a)}\int_0^{\bar t}\rmd t\,t^{\a-1} L(\bar t-t)\,,\label{rli3}
\ee
the Riemann--Liouville integral can be mapped onto a Weyl integral \cite{MR} in the limit $\bar t\to+\infty$. The limit is formal if the Lagrangian $L$ in eq.~\Eq{rli3} is not autonomous. We assume otherwise, so that $\lim_{\bar t\to\infty} L(\bar t-t)\equiv L[q(t),\dot q(t)]$.

This form will be the most convenient for defining a Stieltjes field theory action. When $\bar t\to+\infty$, eq.~\Eq{rli3} is proportional to the usual formula in $\a$ dimensions employed in dimensional regularization. After constructing a ``fractional phase space'' \cite{CSM,Tar1,Tar2,Tar3}, this analogy confirms the interpretation of the order of the fractional integral as the Hausdorff dimension of the underlying fractal \cite{Tar1}.

All the above results in one dimension can be easily generalized to a $D$-dimensional Euclidean space (e.g., \cite{SKM93,Ta06}), thus opening a possibility of applications in spacetime. We entertain the possibility of formulating a scalar field theory with Stieltjes action, for the purpose of controlling its properties in the ultraviolet.\footnote{Introductions on the Lebesgue--Stieltjes integral can be found in \cite{RN,CvB,deb}. A neat geometrical interpretation of the Riemann--Stieltjes one-dimensional integral as the projected ``shadow of a fence'' is given in \cite{Bul88,Pod02}. Projection is a tool sometimes employed to determine the Hausdorff dimension of a fractal \cite{Fal03}.} The scalar field model is interesting in its own right but also as a simple example whereon to work out the physics. After that, we shall explore the gravitational sector. In $D$ dimensions, we consider the action
\be
S=\int_\cM\rmd\vr(x)\, \cL(\phi,\p_\mu\phi)\,,\label{stmes}
\ee
where $\cL$ is the Lagrangian density of the scalar field $\phi(x)$ and 
\be
\rmd \vr(x) = \prod_{\mu=0}^{D-1} f_{(\mu)}(x)\,\rmd x^\mu \label{stme}
\ee
is some multi-dimensional Lebesgue--Stieltjes measure (actually a Lebesgue measure, if $\vr$ is absolutely continuous, which we assume to be the case) generalizing the trivial $D$-dimensional measure $\rmd^{D}x$. We denote with $(\cM,\vr)$ the metric spacetime $\cM$ equipped with measure $\vr$. We shall consider the situation where $\cM$ is a manifold but this may not be the case in general.

Equation \Eq{stmes} resembles a field theory with a dilaton or conformal rescaling $v=\prod_\mu f_{(\mu)}$ of the Minkowski determinant. As in these other models, one will obtain an extra friction term in the equation of motion, although the physics will be radically different both at microscopic and macroscopic level. This is because the measure weight must scale in a certain way, while dilaton solutions in effective actions of string theory typically enjoy much more freedom.
We should stress at least two more reasons why the present model is not just an exotic reformulation of dilaton scenarios.\footnote{These reasons do not forbid the fractal model to admit \emph{also} a ``dilaton-like'' reformulation (see below).} First, the dilaton of string theory couples differently in different sectors, thus leading to a violation of the strong equivalence principle; in our case, the scalar field $v$ is still of geometric origin but appears as a global rescaling. Second, a change in the measure is accompanied by a new definition of functional variations and Dirac distributions, in turn leading to an unfamiliar propagator and the deformation of the Poincar\'e algebra.

If $\vr$ is not invariant under the Lorentz group $SO(D-1,1)$, the equipped manifold is not isotropic even if $\cM$ is the Minkowski flat manifold. If only global Poincar\'e invariance is broken, as it typically happens in fractals, $(\cM,\vr)$ is not homogeneous. Since we wish the Lorentz group $SO(D-1,1)$ to be part of the symmetry group of the action, the Lagrangian density $\cL$ and the $D$ weights $f_{(\mu)}$ must be Lorentz scalars separately. The former can be taken to be the usual scalar field Lagrangian,
\be\label{rlsL}
\cL=-\frac12\p_\mu\phi\p^\mu\phi-V(\phi)\,,
\ee
where $V$ is a potential and contraction of Lorentz indices is done via the Minkowski metric $\eta_{\mu\nu}= (-+\dots+)_{\mu\nu}$. As for the Stieltjes measure, we make the spacetime isotropic choice
\be\label{isot}
f_{(\mu)}=f\,,\qquad \mu=0,1,\dots,D-1\,.
\ee
This should eventually correspond to a fractal in time and space. There are many other \emph{Ans\"atze}, for instance an isotropic nontrivial measure
\be
f_{(0)}=1\,,\qquad f_{(i)}=f\,,\qquad i=1,\dots,D-1\,,
\ee
or an anisotropic measure of the form
\be
f_{(\mu)}= \left\{ \begin{matrix} 1\,,&\quad \mu=0,\dots,i-1\\
f\,,&\quad \mu=i,\dots,D-1\end{matrix}\right..\label{anisot}
\ee
These measures will correspond to different dynamics but, by construction, to the same UV Hausdorff dimension $d_{\rm H}\sim 2$. We take eq.~\Eq{isot} with
\be\label{rlsf}
v\equiv f^{D}\,,\qquad [v]= D(1-\a)\,,
\ee
which generalizes eq.~\Eq{rli3}.\footnote{$v=v(x)$ is a coordinate-dependent Lorentz scalar. An alternative generalization of non-relativistic fractals might have been to choose $v$ to be also metric dependent, e.g., $v=|g_{\mu\nu}\chi^\mu\chi^\nu|^{\frac{D(\a-1)}{2}}$ for some vector $\chi^\mu$. In this case, however, one does not obtain a consistent set of equations of motion.} The scalar field action reads
\be\label{rls}
S=-\int \rmd^{D}x\,v\left[\frac12\p_\mu\phi\p^\mu\phi+V(\phi)\right]\,.
\ee
We now pause and discuss the interpretation of the measure. Classically, one can boost solutions of the equation of motion to a Lorentz frame where $v=v({\bf x})$ (spacelike fractal) or $v=v(t)$ (timelike fractal). These two cases will lead to different classical physics but
at quantum level all configurations should be taken into account, so there is no quantum analogue of space- or timelike fractals.\footnote{We have seen that, approximately, fractional integrals can represent systems living on a certain class of fractal sets. Since we will assume the existence of a nontrivial renormalization group flow entailing integrals of different orders $\a_1,\a_2,\dots$, the complete all-scale picture beyond classical level will not be a fractal with scale-independent Hausdorff dimension but a \emph{multifractal}. For this and the reason stated in the text, the fractal interpretation of fractional integrals is more involved in the quantum theory. To our purposes it is not necessary to stick with it, although we shall do so with a slight abuse of terminology. We shall see that the spectral dimension of the universe changes in a precise, $\a$-dependent way, thus justifying the term ``fractal'' \emph{a posteriori}.} In any case, we shall see that the theory \emph{on the $D\a$-dimensional fractal} is expected to be \emph{dissipative}, i.e., nonunitary. This conclusion is in line with the known results of fractional mechanical systems. Fortunately this will not be a problem because, from the point of view of the manifold $\cM$ with $D$ topological dimensions, energy and momentum are indeed conserved.


\subsection{Renormalization}\label{renor}

The scaling dimension of $\phi$ is
\be
[\phi]=\frac{D\a-2}{2}\,,
\ee
which is zero if, and only if,
\be\label{a2d}
\a=\frac{2}{D}\,.
\ee
Then $\a=1/2$ in four dimensions. This value can change for the other measures defined in eqs.~\Eq{isot} and \Eq{anisot}. In the example \Eq{anisot} with $[f]=1-\a$, one has $\a=2/(D-i)$, and in order for the integral to be properly fractional (rather than multiple) it must be $i\leq D-3$. In four dimensions, there can be at most one ordinary direction. If $i=1$, then $\a=2/3$. $i=0$ corresponds to eq.~\Eq{rlsf}, which we shall adopt from now on.

Let the scalar field potential be polynomial, 
\be
V=\sum_{n=0}^N \s_n\phi^n\,,
\ee
and let $N$ be the highest (positive) power. The coupling $\s_N$ has engineering dimension
\be
[\s_N]=D\a-\frac{N(D\a-2)}{2}\,.
\ee
For the theory to be power-counting renormalizable $[\s_N]\geq 0$, implying
\ba
&&N\leq \frac{2D\a}{D\a-2}\qquad {\rm if}\quad \a>\frac{2}{D}\,,\label{n}\\
&& N\leq+\infty ~~\quad\qquad {\rm if}\quad \a\leq\frac{2}{D}\,.
\ea
When $\a=1$, one gets the standard results $[\phi]=(D-2)/2$, $N\leq 2D/(D-2)$; in four dimensions, the $\phi^4$ theory is renormalizable. In two dimensions, $N$ is unconstrained.

These considerations lead us to try to have the parameters run from an ultraviolet nontrivial fixed point where $\a=2/D$ to an infrared fixed point where, \emph{effectively}, $\a=\a_{\rm IR}$. The dimension of spacetime is well constrained to be 4 from particle physics to cosmological scales and starting at least from the last scattering era \cite{ZS,JY,MS,CO}. Therefore, $\a_{\rm IR}=1$ if $D=4$. To actually realize this particular flow, one should add relevant operators to the action corresponding to terms with trivial measure weight. The total scalar action then is
\be\label{reno}
S=\int\rmd^D x\left[v\cL+M^{D(1-\a)}\tilde\cL\right]\,,
\ee
where $M$ is a constant mass term ($[M]=1$) and $\tilde\cL$ is $\cL$, eq.~\Eq{rlsL}, with all different bare couplings ($\s_n\to\tilde\s_n$). We symbolically represent this modification of the action as
\be\label{mea}
v(x)\to v(x)+M^{D(1-\a)}\,.
\ee
The constant term is anyway expected in the most general Lorentz-invariant definition of the measure weight.

A fractal structure must shortly evolve to a smooth configuration. Oscillations of neutral $B$ mesons can constrain the typical UV mass scale to be larger than about \cite{She09}
\be
M> 300\div 400~ {\rm GeV}.
\ee
Here we do not attempt to place constraints on this scale with other high-energy observations.

As one will see in section \ref{prok}, convergence of the Feynman diagrams is better than in four dimensions, as one can check by looking at the superficial degree of divergence, which is the same as for a $D\a$-dimensional theory \cite{Svo87}. In the case of gravity, in fact, the usual configuration-space results in $2+\epsilon$ dimensions should apply near the UV fixed point \cite{GKT,ChD,Wei79,KN,JJ,KKN,AKNT}.

Needless to say, the above construction and remarks fall short of demonstrating the existence and effectiveness of such a flow, which should be verified by explicit calculations. Our attitude will be to introduce the model and first see its characteristic features and possible advantages, leaving the issue of actual renormalizability for the future.

At any rate, classically the system \emph{will} flow from a lower-dimensional fractal configuration to a smooth $D$-dimensional one. This is clear from the definition \Eq{mea} of the measure weight and its scaling properties when $\a<1$. At small space-time scales, the weight $v\sim |x|^{D(\a-1)}$ dominates over the constant term, while at large scales it is negligible. This is true simply by construction, and independently from renormalization issues.

Therefore, the phenomenological valence of the model is guaranteed, at least. In our framework both the Newton's coupling and the cosmological constant will vary with time already at classical level. On one hand, in minisuperspace models motivated by other approaches to quantum gravity, the running of the couplings can be implemented at the level of the equations of motion, thus obtaining a high-energy ``improved'' dynamics. This strategy is adopted, for instance, in the Planckian cosmology of asymptotically safe gravity thanks to its renormalization properties \cite{BoR1,ReS3,BoR3}. On the other hand, a phenomenological time-varying dimension can be considered for constraining the transition scale from fractal to four-dimensional physics \cite{She09}. The couplings running is then also obtained in fractal-related cosmological toy models with variable dimension \cite{MN}.\footnote{All these scenarios differ in philosophy with respect to \cite{HW}, where the spacetime dimension is promoted to a dynamical field.} 


\section{Scalar field theory}\label{scal}


\subsection{Equation of motion and Hamiltonian} \label{eoh}

The Euler--Lagrange and Hamilton equations of classical mechanical systems with (absolutely continuous) Stieltjes measure have been discussed in \cite{El05a,El05b,FT3} in the one-dimensional case and \cite{ElT,UO,El09a} in many dimensions. The Euler--Lagrange equation of scalar field theory can be found in \cite{El09b}. 

We can easily adapt the same procedure in our case. From now on we consider only the UV part of the action, setting $M=0$. Any result in the infrared can then be obtained by going to the effective limit $\a\to 1$.

The metric space is equipped with a nontrivial measure and caution should be exercised when performing functional variations. For instance, the correct Dirac distribution is
\be\label{stidel}
1=\int \rmd\vr(x)\,\delta_v^{(D)}(x)\, \,,
\ee
as was also noticed in \cite{Svo87}. Invariance of the action under the infinitesimal shift
\be\label{cs}
\phi\to\phi+\delta\phi
\ee
yields the equation of motion (for a generic weight $v$)
\be
0=\frac{\p \cL}{\p\phi}-\left(\frac{\p_\mu v}{v}+\frac{\rmd}{\rmd x^\mu}\right)\frac{\p \cL}{\p(\p_\mu\phi)}\,.
\ee
From eq.~\Eq{rls} we get
\be\label{eom}
\B\phi+\frac{\p_\mu v}{v}\p^\mu\phi-V'=0\,,
\ee
where $\B=\p_\mu\p^\mu$ and a prime denotes differentiation with respect to $\phi$. 
  
The above friction term is characteristic of dissipative systems and one would expect energy not to be conserved. In fact, the Hamiltonian is no longer an integral of motion. Let us define the momentum
\be\label{pi}
\pi_\phi\equiv \frac{\delta S}{\delta\dot\phi}=\dot\phi\,,
\ee
where dots indicate (total) derivatives with respect to time and we have taken eq.~\Eq{stidel} into account.

Defining the Lagrangian
\be\label{lag}
L\equiv \int \rmd{\bf x}\,v\cL\,,
\ee
the Hamiltonian is
\ba
H&\equiv& \int\rmd{\bf x}\,v\,(\pi_\phi\dot\phi)-L\label{ham}\\
&=& \int\rmd{\bf x}\,v\left(\tfrac12\pi_\phi^2+\tfrac12\p_i\phi\p^i\phi+V\right)\nonumber\\
&\equiv& \int\rmd{\bf x}\,v\, \cH\,,
\ea
where $\rmd{\bf x}=\rmd x^1\dots\rmd x^{D-1}$. The definition of the equal-time Poisson brackets (time dependence implicit)
\be
\{A({\bf x}),B({\bf x'})\}_v\equiv \int \rmd\vr({\bf y})\, \left[\frac{\delta A({\bf x})}{\delta\phi({\bf y})}\frac{\delta B({\bf x'})}{\delta\pi_\phi({\bf y})}-\frac{\delta A({\bf x})}{\delta\pi_\phi({\bf y})}\frac{\delta B({\bf x'})}{\delta\phi({\bf y})}\right]
\ee
yields the Hamilton equations
\ba
\dot\phi     &=& \{\phi,H\}_v\,,\label{dp}\\ 
\dot\pi_\phi &=& \{\pi_\phi,H\}_v-\frac{\dot v}{v}\pi_\phi\,,\label{dis}
\ea
equivalent to eqs.~\Eq{pi} and \Eq{eom}, respectively. Therefore, time evolution of an observable $O(\phi,\pi_\phi,x)$ is
\be\label{Ot}
\dot O=\p_t O+\{O,H\}_v-\frac{\dot v}{v}\pi_\phi\frac{\p O}{\p\pi_\phi}\,.
\ee
Equations \Eq{dis} and \Eq{Ot} signal dissipation. Nonconservation of the Hamiltonian is expected from the definition of the Lagrangian \Eq{lag}. The measure factor $v$ is both time and space dependent, so it was not possible to factorize it in order to write the action as the correct Stieltjes time integral of $L$. However, one can exploit Lorentz invariance and pick a frame where $v=v(|{\bf x}|)$. In this frame $H$ would be restored as the generator of time translations; for a timelike fractal it is not possible to pick this frame and there is always energy dissipation. Conversely, physical momentum will be dissipated in a spacelike fractal.

Is there a problem with that? Whatever the choice of classical fractal model, one would have to face the issue of unitarity at quantum level. Moreover, we need a physical interpretation of dissipation. It turns out that the latter helps to address the above concern.

Consider a $(D-1)$-dimensional box of size $l$ and spatial volume $l^{D-1}$. At the scale $l$, particles live effectively in $D\a$ spacetime dimensions. If $\a=1$, they occupy the whole phase space in the box. Otherwise, they must dissipate energy, since the energy of the configuration filling the entire topological volume is different from that of a configuration limited to the effective $D\a$-dimensional world. The total energy of the system $\cE$ in $D$ topological dimensions is conserved, but the energy $H$ measured by a $D\a$-dimensional observer is not (eq.~\Eq{Ot} with $O=\cH v$ and integrated in space):
\ba
\dot H &=&\p_t H-\int \rmd{\bf x}\,\dot v \pi^2_\phi\nonumber\\
       &=& -\int\rmd{\bf x}\, \dot v \cL\,.\label{Ht}
\ea
In fact, a conserved quantity is
\ba
\cE(t)     &=& H(t)+\Lambda(t)\,,\\
\Lambda(t) &=& \int^t\rmd t\int\rmd{\bf x}\,\dot v\cL\,,\label{Lam}
\ea
which we are going to obtain also from Noether's theorem (see \cite{FT3,DV} for a computation in classical mechanics). It should be easy to see that $\Lambda$ can be regarded as the \emph{complementary function} of initialized fractional calculus \cite{Das08,LH00,NLH,LH08}. Physically, it is a running cosmological constant of purely geometric origin. For this reason, dissipation might eventually prove to be an asset rather than a liability of the theory.

Unlike standard scalar field theory, the Noether current associated with the usual Lagrangian continuous symmetries is not covariantly conserved. On the other hand, one can easily find generalized conserved currents. Take a generic infinitesimal transformation of the field, eq.~\Eq{cs}, and coordinates, $x^\mu\to x^\mu+\delta x^\mu$. We consider symmetries of the autonomous Lagrangian density $\cL$ and define ``quasi invariance'' of the action as done in \cite{FT3,El09b,DV}. Then $\delta\cL$ is a total divergence,
\be
\cL\to \cL+\frac{\rmd{\cal J}^\mu}{\rmd x^\mu}\,.
\ee
Combining this equation with
\ba
\delta\cL &=&\frac{\p\cL}{\p\phi}\delta\phi+\frac{\p\cL}{\p(\p_\mu\phi)}\delta(\p_\mu\phi)\nonumber\\
&=&\frac{\p\cL}{\p\phi}\delta\phi+\frac{\p\cL}{\p(\p_\mu\phi)}\p_\mu(\delta\phi)\nonumber\\
&=&\left(\frac{\p\cL}{\p\phi}-\frac{\rmd}{\rmd x^\mu}\frac{\p \cL}{\p\p_\mu\phi}\right)\delta\phi+\frac{\rmd}{\rmd x^\mu}\left(\frac{\p \cL}{\p\p_\mu\phi}\delta\phi\right)\,,\nonumber\\
\ea
one obtains on shell
\be
-\frac{\rmd}{\rmd x^\mu}\left[v\left(\cJ^\mu-\frac{\p \cL}{\p\p_\mu\phi}\delta\phi\right)\right]+\cJ^\mu\p_\mu v=0\,.\label{no0}
\ee
Choose a coordinate translation 
\be\label{tran}
{x^\mu}'=x^\mu+\delta x^\mu\,,\qquad \delta x^\nu=-a^\nu\,,
\ee
so that $\delta\phi=a^\nu\p_\nu\phi$ and $\delta\cL=a^\nu \rmd\cL/\rmd x^\nu$. Then
\be
-\frac{\rmd}{\rmd x^\mu}\left(v T^\mu_{\ \nu}\right)+\cL\p_\nu v=0\,,\label{noe1}
\ee
where
\be
T^\mu_{\ \nu}\equiv \delta^\mu_\nu\cL+\p^\mu\phi\p_\nu\phi
\ee
is the usual energy-momentum tensor. Integrating eq.~\Eq{noe1} in space, one gets
\be
\dot P_{\nu}+\int\rmd{\bf x}\,\p_\nu v\,\cL =0\,,\label{noe2}
\ee
where
\be
P_\nu\equiv-\int\rmd{\bf x}\, v\, T^0_{\ \nu}\,.
\ee
The $\nu=0$ component yields $P_0=H$ and eq.~\Eq{Ht} follows suit. The $\nu=i$ component gives the physical momentum  
\be
P_i=\int\rmd{\bf x}\, v\, \pi_\phi\p_i\phi\,,
\ee
and its conservation law
\be\label{Pt}
\dot P_i+\dot \Lambda_i\equiv\dot P_i+\int\rmd{\bf x}\,\p_iv\,\cL=0\,.
\ee
$P_i$ generates spatial translations in the field but not in its conjugate momentum, since the covariant counterparts of eqs.~\Eq{dp} and \Eq{dis} are
\ba
\p_\mu\phi     &=& \{\phi,P_\mu\}_v\,,\label{dpi}\\
\p_\mu\pi_\phi &=& \{\pi_\phi,P_\mu\}_v-\frac{\p_\mu v}{v}\pi_\phi\,.\label{disi}
\ea
The covariant version of eq.~\Eq{Ot} is
\be\label{Oi}
\frac{\rmd O}{\rmd x^\mu}=\p_\mu O+\{O,P_\mu\}_v-\frac{\p_\mu v}{v}\pi_\phi\frac{\p O}{\p\pi_\phi}\,,
\ee
yielding 
\be\label{Pmu}
\frac{\rmd P_\nu}{\rmd x^\mu}=\{P_\nu,P_\mu\}_v-\delta^0_\nu\int\rmd{\bf x}\,\p_\mu v\,\cL\,.
\ee
The $(\mu,\nu)$ components of this equation offer consistency checks and new commutation relations. Component $(0,0)$ corresponds to eq.~\Eq{Ht}, while $(i,j)$ gives
\be
\{P_i,P_j\}_v=0\,.
\ee
Components $(0,i)$ and $(i,0)$ and eq.~\Eq{Pt} are consistent among each other if, and only if,
\be
\{H,P_i\}_v=\int\rmd{\bf x}\,\p_i v\,\cL\,,
\ee
finally showing that the Poincar\'e algebra is now noncommutative unless $v$ is only time dependent (timelike fractal). One can check that also the Lorentz algebra is deformed. Consider the Noether current associated with boost/rotation transformations
\ba
\delta\phi &=& a_{\nu\s}(x^\nu\p^\s-x^\s\p^\nu)\phi\,,\\
\delta\cL &=& a_{\nu\s}(x^\nu\p^\s-x^\s\p^\nu)\cL=a_{\nu\s}\p_\mu(g^{\mu\s}x^\nu\cL-g^{\mu\nu}x^\s\cL)=\p_\mu\cJ^\mu\,.
\ea
Substituting in eq.~\Eq{no0}, we get
\be
-\frac{\rmd}{\rmd x^\mu}\left(v M^{\mu\nu\s}\right)+\cL(x^\nu\p^\s v-x^\s\p^\nu v)=0\,,\label{noelo}
\ee
where
\be
M^{\mu\nu\s}\equiv x^\nu T^{\mu\s}-x^\s T^{\mu\nu}\,.
\ee
The algebra of the Lorentz generators $J^{\nu\s}\equiv\int \rmd{\bf x} v M^{0\nu\s}$, which includes the Lorentz boosts $K^i\equiv J^{0i}$ and the angular momenta $L^i\equiv \tfrac12\epsilon^i_{\ jk} J^{jk}$ ($\epsilon^i_{\ jk}$ is the Levi-Civita symbol), is deformed because of the nontrivial weight $v$.

We have ended up with a classical field theory living on a $D\a$-dimensional fractal breaking Poincar\'e invariance. Depending whether the fractal is ``timelike'' or ``spacelike,'' there will be a momentum or energy transfer between the world-fractal and the $D$-dimensional embedding.

The $D$-dimensional side of the picture can be actually made more precise. So far we have interpreted the function $v$ in the action \Eq{rls} as (the derivative of) a Stieltjes measure defined on a fractal of Hausdorff dimension $D\a$. Of course one can regard it as a ``dilaton'' field coupled with the Lagrangian density $\cL$ living on a $D$-dimensional manifold. Then it is natural to consider the usual $\delta$ of Dirac,
\be\label{del}
1=\int\rmd^D x\, \delta^{(D)}(x)\,.
\ee
Consequently, the momentum conjugate to the field $\phi$ is
\be
\pi_v= v\pi_\phi\,.
\ee
It is easy to convince oneself that the Poisson brackets are the usual
\be
\left\{\,\cdot\,\vphantom{H},\,\cdot\,\vphantom{H}\right\}_1\,,
\ee
and that all $v$ dependence disappears in the $D$-momentum
\be
P_\mu=P_\mu(\phi,\pi_v)\,.
\ee
Poincar\'e (and Lorentz) invariance is preserved. This completes the proof that, at least at classical level, dissipation occurs relatively between parts of a \emph{conservative} system. Quantization would follow through, although an UV observer would experience an effective probability flow from or into his world-fractal (see also \cite{Pel91}).

To summarize, because the universe is associated with a ``fractal'' structure one typically expect to have breaking of Lorentz invariance at small scales. The geometry of the problem is not standard and this modifies the usual definition of Dirac distribution and Poisson brackets. As a result, from the point of view of an observer living in the fractal and measuring geometry with weight $v$, Lorentz and translation invariance are broken inasmuch as the action itself is Lorentz and Poincar\'e invariant, but the algebra of the Poincar\'e group is deformed. In other words, when talking about fractals embedded in Minkowski spacetime we mean \emph{the geometries defined by the deformed Poincar\'e group}. However, from the point of view of the ambient $D$-dimensional manifold Poisson brackets and functional variations no longer feature the nontrivial measure weight, which is now regarded as an independent matter field rescaling. In that case, the full Poincar\'e group is preserved. On the other hand, the properties of \emph{nonrelativistic} fractals are more intuitive: they break translation invariance because the measure weight introduces explicit coordinate dependence and the system is not autonomous.


\subsection{Propagator: configuration space}\label{proc}

The theory is Lorentz invariant, ghost free and causal at all scales. We can check this explicitly by computing the causal propagator $G(x)$, which is (proportional to) the propagator in two and $D$ dimensions at the UV and IR fixed points, respectively.

As usual, we define the vacuum-to-vacuum amplitude (partition function)
\be\label{parf}
Z[J]=\int[\rmd\phi]\,\rme^{\rmi\int \rmd\vr (\cL+\phi J)}\,,
\ee
where $J$ is a source and we have already integrated out momenta. Integration by parts in the exponent allows us to write the Lagrangian density for a free field as
\be
\cL=\frac12\phi\left(\B+\frac{\p_\mu v}{v}\p^\mu-m^2\right)\phi\equiv \frac12\phi \cC \phi\,.
\ee
The propagator is the Green function solving
\be\label{gfe}
\cC G(x)=\delta^{(D)}_v(x)\,.
\ee
By virtue of Lorentz covariance, the Green function $G$ must depend only on the Lorentz interval
\be
s^2=x_\mu x^\mu=x_ix^i-t^2\,,
\ee
where $t=x^0$ and $i=1,\dots,D-1$. In particular, $v=v(s)$ with the correct scaling property is $v(s)\sim |s|^{D(\a-1)}$. This definition guarantees reality of the measure and avoids problems with unitarity (in particular, the action is real).

One might be worried that the measure blows up on the light cone, but this is an integrable singularity: the check that $\int_A\rmd^D x\, v<\infty$ is done on a compact set $A$ (for instance, a $D$-ball of radius $R$) and in Euclidean signature. Below we will also see that there is nothing pathological in the propagator on the light-cone, even if $v$ is singular in $s=0$. Measures describing fractals may be very irregular and it would be worth investigating the physical interpretation of their singularities, especially in Lorentzian signature. For the time being, we notice that one can also define the measure weight to be
\be
v(s)=\left\{ \begin{matrix} |s|^{D(\a-1)}\,,&\quad s\neq 0\\
1\,,&\quad s=0\end{matrix}\right.\,.\label{rlsf2}
\ee
This definition may be in contrast with the original assumption that the measure be absolutely continuous. This would mean that the simplified model with an overall measure weight does not come from a most general fractal model with Lebesgue--Stieltjes measure $\vr$: in other words, $\rmd\vr(x)\neq v(x)\,\rmd^Dx$. This is not a problem for two reasons. On one hand, the simple model with measure weight $v$ is still able to capture much of the physics of the general Stieltjes model, by virtue of the scaling argument; as a matter of fact, one could even take measures which are Lebesgue--Stieltjes only asymptotically, and yet obtain a modelization of a fractal quantum field theory in certain regimes. On the other hand, one can devise other measure profiles with the same scaling properties and regular behaviour. As an example, instead of eq.~\Eq{rlsf2} one could take
\be\label{posal}
v(s)=\frac{1}{2\ell^{D(1-\a)}}+\frac{1}{|s|^{D(1-\a)}+2\ell^{D(1-\a)}},
\ee
where $\ell=1/M$. At small $s$ (near the light cone) or large space/time scales, $v\to {\rm const}$; at intermediate scales, $v$ has the power-law behaviour which we will assume from now on.

Without risk of confusion, we use the symbol $\p$ to denote total derivatives. Noting that
\be
\p_\mu =\frac{x_\mu}{s}\p_s\,,\qquad \B=\p_s^2+\frac{D-1}{s}\p_s\,,
\ee
the inhomogeneous equation \Eq{gfe} reads
\be\label{gree}
\left(\p_s^2+\frac{D\a-1}{s}\p_s-m^2\right)G(s)=\delta^{(D)}_v(x)\,,
\ee
where we added a mass term, $[m]=1$, $m^2>0$. We first consider the Euclidean propagator and denote with $r=\sqrt{x_i x^i+t^2}$ the Wick-rotated Lorentz invariant.

In the massless case, the solution of the homogeneous equation is
\be\label{Gm0}
G= C (r^2)^{1-\frac{D\a}2}\,,
\ee
where $C$ is a normalization constant. The right-hand side of eq.~\Eq{gree} is not the $\delta$ defined in radial coordinates. To find the latter, one notices that
\ba
1 &=& \int \rmd^Dx\, v\delta^{(D)}_v(x)\nonumber\\
  &=& \Om_D \int \rmd r\, v\, r^{D-1}\delta^{(D)}_v(x)\nonumber\\
  &=& \int \rmd r\, \delta(r)\,,\nonumber
\ea
where $\Om_D=2\pi^{D/2}/\Gamma(D/2)$ is the volume of the unit $D$-ball. Therefore,
\be
\int\rmd\vr(x)\,\delta^{(D)}_v(x)=\int\rmd\vr(x)\left[\frac{r^{1-D}}{\Om_D v(r)}\delta(r)\right]\,.
\ee
Hence, to find the propagator also for $r=0$ one can take some test function $\vp$ and compute
\be
\Om_D\langle \cK G,\vp\rangle=\lim_{\epsilon\to 0} \Om_D\int_\epsilon^{+\infty} \rmd r\,\cK G(r) \vp(r)\,,
\ee
where
\ba
\cK &=& v(r) r^{D-1} \cC\nonumber\\
    &=& \p_r\left(r^{D\a-1}\p_r\right)-r^{D\a-1}m^2\,.
\ea
Therefore,
\ba
\Om_D\langle \cK G,\vp\rangle &=& \Om_D\langle G,\cK\vp\rangle\nonumber\\
&=& \lim_{\epsilon\to 0} \Om_D\int_\epsilon^{+\infty} \rmd r\, G(r) \p_r\left(r^{D\a-1}\p_r\vp\right)\nonumber\\
&=& C\Om_D(2-D\a)\vp(0)\,,\nonumber
\ea
where we have used eq.~\Eq{Gm0} and integrated by parts once (boundary terms vanish). The last line must be equal to $\langle \delta,\vp\rangle$, thus fixing $C$. Then, the Green function for $m=0$ reads
\be
G(r) =\frac{1}{\Om_D(2-D\a)}(r^2)^{1-\frac{D\a}2}\,.\label{prop0}
\ee
This result enjoys several consistency checks. When $\a=1$, it is the usual Green function $G_D$ for the Laplacian in $D$ dimensions with standard Lebesgue measure:
\be\nonumber
\lim_{\a\to 1}G(r)=G_D(r)=-\frac{\Gamma\left(\tfrac{D}{2}-1\right)}{4\pi^{D/2}} (r^2)^{1-\frac{D}2}\,.
\ee
In the UV limit $\a\to 2/D$, one can expand eq.~\Eq{prop0} as $r^\epsilon/\epsilon= 1/\epsilon+\ln r+O(\epsilon)$. Up to a divergent constant, this is the logarithmic propagator in two dimensions, rescaled with a volume ratio:
\be\label{logpr}
G_*(r)\equiv\lim_{\a\to 2/D}G(r)=\frac{\Om_2}{\Om_D}\,G_2(r)=\frac{1}{\Om_D}\ln r\,.
\ee
As a side remark, notice that $G$ can be written as
\be\label{mle}
G(r)\propto \frac{\rmd^{-\b}}{\rmd r^{-\b}} G_D(r)\,,
\ee
where
\be\label{beta}
\b\equiv D(1-\a)\,,
\ee
and the fractional derivative can be defined as the Liouville derivative:
\be\label{liod}
\frac{\rmd^{-\b}}{\rmd r^{-\b}}r^\g = \frac{\Gamma(\g+1)}{\Gamma(\g+\b+1)}r^{\g+\b}\,.
\ee
The order of the derivative is negative, so that it is actually a fractional integral with clear meaning: Starting from the problem $\B G_D=\delta^{(D)}$ and inserting (heuristically) the identity $\p^\b\p^{-\b}$, it replaces the second-order operator on $G_D$ with a fractional differentiation of order $2+\b$ on $G$. We can conclude that
\be\label{dua2}
G(r) \propto G^{(1+\b/2)}(r)\,,
\ee
i.e., the (massless) propagator is proportional to the Green function solving the pseudodifferential equation
\be\label{frag}
\B^{1+\b/2} G^{(1+\b/2)}=\delta^{(D)}\,,
\ee
which was calculated and discussed in \cite{BGG,Gia91,BGO,BG,doA92,BBOR} and reads
\be\label{dua1}
G^{(1+\b/2)}\propto  (s^2+\rmi\ve)^{1+\frac{\b-D}2}\,.
\ee
Indeed, after Wick rotation eqs.~\Eq{Gm0} and \Eq{dua1} agree up to the normalization. In this sense, in configuration space our field theory on a fractal is equivalent to a certain class of nonlocal models represented by eq.~\Eq{frag} \cite{doA92,BBOR,BOR}.

We now consider the massive case (Helmholtz equation). The solution of the homogeneous equation $\cC G=0$ is
\be\label{Gm}
G(r) = \left(\frac{m}{r}\right)^{\frac{D\a}{2}-1}\left[C_1 K_{\frac{D\a}{2}-1}(mr)+C_2 I_{\frac{D\a}{2}-1}(mr)\right]\,,
\ee
where $C_{1,2}$ are constants and $K$ and $I$ are the modified Bessel functions. Since for small $m$ the solution must agree with the massless case \Eq{Gm0}, we can set $C_2=0$. In fact, this is true only for $\a\neq 2/D$, as one can see from the asymptotic formul\ae\
\ba
K_\nu(z) &\approx&  \left\{ \begin{matrix}
  - \ln \left(\frac{z}2\right) - \gamma   & ~~\mbox{if } \nu=0 \\ \\
  \frac{\Gamma(\nu)}{2} \left(\frac{2}{z}\right)^\nu & ~~\mbox{if } \nu > 0 
\end{matrix} \right. ,\label{asiK}\\
I_\nu(z) &\approx& \frac{1}{\Gamma(\nu+1)}\left(\frac{z}{2}\right)^\nu\,,\label{asiI}
\ea
where $\gamma$ is the Euler--Mascheroni constant. We shall discuss the case $\a=2/D$ separately.\footnote{The comparison with field theory with fractional powers of the d'Alembertian holds also when $m\neq 0$. Recalling that 
\be
\left(\frac1z \frac{\rmd}{\rmd z}\right)^n [z^{-\nu} K_\nu(z)]=(-1)^n z^{-\nu-n} K_{\nu+n}(z)\,,\qquad n\in\mathbb{N}\,,\nonumber
\ee
one can continue this formula to any $n$ and write the propagator as
\be\nonumber
G(r)\propto \left(\frac1r \frac{\rmd}{\rmd r}\right)^{-\frac{\b}{2}} G_D(r)\,.
\ee}

To find the solution of the inhomogeneous equation, one exploits the fact that the mass term does not contribute near the origin. Expanding eq.~\Eq{Gm} at $mr\sim 0$  when $D\a> 2$ ($C_2=0$),
\be\nonumber
G(r) \sim C_1 2^{\frac{D\a}{2}-2} \Gamma\left(\tfrac{D\a}{2}-1\right)(r^2)^{1-\frac{D\a}{2}}\,,
\ee
which must coincide with eq.~\Eq{prop0}. This fixes the coefficient $C_1$ and the propagator reads
\be\label{propm}
G(r)  =-\frac{1}{2\pi^{\frac{D}{2}}}\frac{\Gamma\left(\tfrac{D}2\right)}{\Gamma\left(\tfrac{D\a}2\right)}\left(\frac{m}{2r}\right)^{\frac{D\a}{2}-1} K_{\frac{D\a}{2}-1}(mr)\,,
\ee
in agreement with the Helmholtz propagator ($\a=1$).

Here an important remark is mandatory. In usual quantum field theory, the propagator $G(r)$ is defined up to an immaterial constant $C$. By ``immaterial'' we obviously mean that $\B[G(r)+C]=\B G(r)$. This is true in any dimension $D$ and regardless the functional form of $G(r)$, being it just a property of ordinary differentiation, $\B C=0$. Our model is still characterized by ordinary differential operators, so the set
\be\label{eqc}
\left\{G(r)+C~\big|~C={\rm const}\right\}
\ee
is an equivalence class defining the propagator. However, the nontrivial measure $\vr$ will associate elements of this equivalence class with different Fourier transforms in momentum space. In the critical case $\a=2/D$, the difference will be in unusual terms
\be\nonumber
B(C_1,C_2,C)\frac{1}{k^2}
\ee
which resemble massless poles. If the theory is well defined, these terms will have to correspond to generalized Dirac distributions, and \emph{not} to particle modes with arbitrarily chosen residue $B(C_1,C_2,C)$. Keeping this in mind will prevent us to fall into a false paradox when calculating the propagator in momentum space.

Taking this issue on board, the case $\a=2/D$, $m\neq 0$ is straightforward: it is safe to just set $\a=2/D$ in the noncritical propagator \Eq{propm},
\be\label{props2}
G_*(r)=-\frac{1}{\Om_D} K_0(mr)\,.
\ee

Now to the Lorentzian theory. The fractal field models of \cite{Svo87} and \cite{Ey89a} both meet the Osterwalder--Schrader conditions. This encourages the expectation that a generic field theory on a Euclidean fractal, if well defined, should admit an analytic continuation to a theory in Lorentz spacetime. In fact, the Euclidean partition function eq.~\Eq{parf} is a Schwinger function endowed with all the properties required by the Osterwalder--Schrader theorem: it is analytic, symmetric under the permutation of arguments, Euclidean covariant, and satisfies cluster decomposition and reflection positivity. Consequently, we can analytically continue the Helmholtz propagator \Eq{propm} to the Klein--Gordon propagator according to the prescriptions: (i) multiply $G$ times the imaginary unit $\rmi$, due to Wick rotation of the time direction; (ii) replace $r^2$ with $s^2+\rmi\ve$, where the positive sign of the extra infinitesimal term corresponds to the \emph{causal} Feynman propagator. Summarizing,
\be
G(s)  =-\frac{\rmi}{2^{\frac{D\a}{2}}\pi^{\frac{D}{2}}} \frac{\Gamma\left(\tfrac{D}2\right)}{\Gamma\left(\tfrac{D\a}2\right)} \left(\frac{m^2}{s^2+\rmi\ve}\right)^{\frac{D\a}{4}-\frac12} K_{\frac{D\a}{2}-1}\left(m\sqrt{s^2+\rmi\ve}\right)\,,\qquad s>0\,,\label{propmm}
\ee
while the massless propagator is
\be\label{promima}
G(s) =\frac{\rmi}{\Om_D(2-D\a)}(s^2+\rmi\ve)^{1-\frac{D\a}2}\,,
\ee
in accordance with eq.~\Eq{dua1}. Equations \Eq{logpr} and \Eq{props2} are continued similarly.

The propagator for timelike intervals is just the analytic continuation of the former. In the massive case, it is proportional to the Hankel function of the first kind $H^{(1)}_{D\a/2-1}$. One last thing to check is what happens on the light cone. The propagator for $D\a=2$, eq.~\Eq{logpr}, is (proportional to) the usual propagator in two dimensions, so nothing special occurs. 
Taking instead the definition \Eq{rlsf2}, setting $\a=1$ in eq.~\Eq{promima} (contribution of $m$ negligible), for even $D=4,6,\dots$ one obtains (\cite{GeS}, p.~94)
\ba
G(s) &\propto& \frac1{(s^2+\rmi\ve)^{\frac{D}2-1}}\nonumber\\
&=&\frac{(-\p_{s^2})^{\frac{D}2-2}}{(D/2-2)!}\left[{\rm PV}\left(\frac1{s^2}\right)
-\rmi \pi\delta(s^2)\right]\,,\nonumber\\\label{soc}
\ea
where PV denotes the principal value. In $D=4$ this reduces to the Plemelji--Sokhotski formula. Translated into momentum space, the $\delta$ states, as usual, that massless particles propagate at the speed of light on the light cone (Huygens' principle \cite{Gia91,BG}).


\subsection{Propagator: momentum space}\label{prok} 

Since $G$ has been argued to be proportional, in configuration space, to the Green function of another well-known problem (the functional inverse of a fractional power of the d'Alembertian), we can already guess its pole structure in momentum space: in general, it will exhibit a branch cut with branch point at $k^2=-m^2$.

To calculate the propagator in momentum space, we can start from the Euclidean one and then analytically continue the result as usual. In the Lorentzian propagators the substitution $k^2\to |{\bf k}|^2-(k^0)^2-\rmi\ve$ is understood.

In the presence of a Lebesgue--Stieltjes measure, the Fourier transform must be modified so that it is consistent with the definition of the Dirac distribution eq.~\Eq{stidel}. The Fourier--Stieltjes transform $F_v$ of a function $G(x)$ and its inverse are defined as \cite{Svo87}
\ba
\tilde G(k) &=& \int\rmd\vr(x)\, G(x)\,\rme^{-\rmi k\cdot x}\equiv F_v[G(x)]\,,\\
G(x) &=& \frac{1}{(2\pi)^D}\int\rmd\vr(k)\, \tilde G(k)\,\rme^{\rmi k\cdot x}\,.\label{fst}
\ea
The measure in eq.~\Eq{fst} is such that momentum and configuration space have the same dimensionality. The Fourier--Stieltjes transform of a function $G$ is the Fourier transform of $vG$:
\be\label{ffv}
F_v[G] = F[v G] = F[r^{D(\a-1)}G(r)]\,.
\ee
In particular, the Fourier--Stieltjes transform of $\delta_v^{(D)}$ is 1. When $\a\neq 1$ one has
\ba
F_v[1] &=& F[r^{D(\a-1)}]\nonumber\\
&=&2^{D\a}\pi^{D/2}\frac{\Gamma\left(\tfrac{D\a}{2}\right)}{\Gamma\left[\tfrac{D(1-\a)}{2}\right]}\frac{1}{k^{D\a}}\nonumber\\
&=&(2\pi)^D\delta_v^{(D)}(k)\,,\label{delk}
\ea
so $\delta_v^{(D)}$, the source in eq.~\Eq{gfe}, is a power-law distribution.

Equation \Eq{ffv} tells us the form of the massless propagator when $\a\neq 2/D$ (transform of eq.~\Eq{prop0})  \cite{GeS}: 
\ba
\tilde G(k) &=& \frac{1}{\Om_D(2-D\a)} F[r^{D(\a-1)}r^{2-D\a}]\nonumber\\
&=& -\frac{D-2}{D\a-2}\frac{1}{k^2}\,.\label{Gk0}
\ea
Notice that:
\begin{itemize}
\item The (Lorentzian) propagator has a $k^2=0$ pole in the $({\rm Re}\,k^0,{\rm Im}\,k^0)$ plane.
From the world-fractal point of view (that is to say, looking at the pole structure of $\tilde G(k)$ rather than of $v(k)\tilde G(k)$), the spectrum has the usual support at $k^2=0$.
\item When $\a=1$, one obtains $\tilde G(k)=-1/k^2$ and the free wave solution.
\item For general $\a>2/D$, the sign of the residue is always negative, which ensures the absence of ghosts. However, its value is not 1 but given by a geometric factor. This is expected as the effective theory in the world-fractal is not unitary and some probability is exchanged with the $D$-dimensional topological bulk.
\end{itemize}
In the critical case $\a=2/D$ eq.~\Eq{Gk0} is ill-defined; this is not a problem, since one should start from eq.~\Eq{logpr}:
\ba
\tilde G_*(k) &=& \frac{1}{\Om_D}F[r^{2-D}\ln r]\nonumber\\
&\ \stackrel{?}{=}\ & \left(\frac{D}{2}-1\right)\frac{B-\ln k^2}{k^2}\,,\label{Gk2}
\ea
where $B=\psi(D/2-1)-\gamma+\ln4$ and $\psi$ is the digamma function. Some remarks:
\begin{itemize}
\item For $D\neq 2$, the Lorentzian propagator has a branch point at $k^2=0$.
\item In the limit $D\to2$, the propagator is $G\sim -1/k^2$.
\item The term $B/k^2$ does not represent a particle mode. In fact, it is nothing but the fractal Dirac distribution eq.~\Eq{delk} when $\a=2/D$. In other words, the Fourier--Stieltjes transform of the equivalence class \Eq{eqc} is unique up to a $\delta_v^{(D)}(k)$ term. The spectrum consists in a quasiparticle continuum of modes with momentum $k^2\leq 0$.
\end{itemize}
In eq.~\Eq{eqc} there exists a particular element such that $B=0$ identically, which we typically call \emph{the} propagator. Equation \Eq{Gk2} is then replaced by the unambiguous expression
\be
\tilde G_*(k) = -\left(\frac{D}{2}-1\right)\frac{\ln k^2}{k^2}\,.\label{Gk22}
\ee
The massive case is slightly more complicated and as an exercise we will calculate it explicitly. The transform of the propagator in radial coordinates is
\be\nonumber
\tilde G(k) =  \int\rmd \Om_D \rmd r\,r^{D-1}\, v(r) G(r)\,\rme^{-\rmi k\cdot x}\,.
\ee
The integrand is not radial but one can choose a frame where $k_\mu x^\mu=-kr\cos\theta$, $k\equiv |k^\mu|$, and the angular integral reads
\ba
\int\rmd\Om_D\,\rme^{-\rmi k\cdot x} &=& \Om_{D-1}\int_0^\pi\rmd\theta (\sin\theta)^{D-2}\rme^{\rmi k r\cos\theta}\nonumber\\
&=& \Om_{D-1} \sqrt{\pi}\,\Gamma\left(\frac{D-1}{2}\right) \left(\frac{2}{kr}\right)^{\frac{D}2-1}J_{\frac{D}2-1}(kr)\,.\nonumber
\ea
In the last line we used formula 3.915.5 of \cite{GR}. 
Then,
\be\label{propkgen}
\tilde G(k) = \Gamma\left(\frac{D}{2}\right)\left(\frac{2}{k}\right)^{\frac{D}2-1}\int_0^{+\infty}\rmd r\,r^{D\a-\frac{D}2}[\Om_DG(r)]J_{\frac{D}2-1}(kr)\,.
\ee

Now we take the massive propagator \Eq{propm} ($\a\neq 2/D$):
\ba
\tilde G(k)&=& -\frac{\Gamma\left(\tfrac{D}{2}\right)}{\Gamma\left(\tfrac{D\a}2\right)}\left(\frac{2}{k}\right)^{\frac{D}2-1}\left(\frac{m}{2}\right)^{\frac{D\a}{2}-1}\nonumber\\
&&\times \int_0^{+\infty}\rmd r\,r^{\frac{D(\a-1)}2+1} K_{\frac{D\a}{2}-1}(mr)J_{\frac{D}2-1}(kr)\nonumber\\
&=& -\frac{F\left(\tfrac{D\a}{2},1;\tfrac{D}2;-\frac{k^2}{m^2}\right)}{m^2}\,,\label{Gkm}
\ea
where we used formula 6.576.3 of \cite{GR} 
and $F$ is the hypergeometric function
\be
F(a,b;c;z)=\sum_{n=0}^\infty\frac{\Gamma(a+n)\Gamma(b+n)}{\Gamma(a)\Gamma(b)}\frac{\Gamma(c)}{\Gamma(c+n)}\frac{z^n}{n!}\,.
\ee
When $\a=1$, one obtains the usual massive propagator in $D$ dimensions (formula 9.121.1 of \cite{GR}): 
\be
\tilde G(k)=-\frac{1}{k^2+m^2}\,.
\ee
In the limit $m\to 0$ (which does commute with the analytic continuation of eq.~\Eq{Gkm}), one recovers eq.~\Eq{Gk0} up to a term $O(m^{D\a-2})$, which is negligible by virtue of the no-ghost condition $\a>2/D$.

When $\a=2/D$, the transform of the critical propagator eq.~\Eq{props2} is
\ba
\tilde G_*(k)&=& -\frac{F\left(1,1;\tfrac{D}2;-\frac{k^2}{m^2}\right)}{m^2}\label{Gkmcr}\\
&=& -\left(\frac{D}{2}-1\right)\frac{\ln k^2}{k^2}+O(m^2\ln m^2)\,,\label{Gk0cr} 
\ea
where in the second line we have dropped $O(k^{-2})$ terms and considered the massless limit for comparison with eq.~\Eq{Gk22}. When $D=4$, by virtue of formula 9.121.6 of \cite{GR}, eq.~\Eq{Gkmcr} gives exactly (i.e., no $\delta_v$ terms)
\be\label{res1}
\tilde G_*(k)= -\frac{1}{k^2}\ln\left(1+\frac{k^2}{m^2}\right)\,,
\ee
which is precisely the generalization of eq.~\Eq{Gk22}. $G_*(k)$ has a branch cut for ${\rm Re}\,k^0<-\sqrt{|{\bf k}|^2+m^2}$. Therefore, the spectrum of the theory has a continuum of modes with rest mass $\geq m$.

Interestingly, the convergence domain of eq.~\Eq{Gkm} gives a bound on the topological dimension $D$. The propagator is a hypergeometric series with convergence in the unit circle $|z|<1$, where $z=-k^2/m^2$. Let
\be
p\equiv \frac{D\a}{2}-\frac{D}{2}+1\,.
\ee
On the unit circle $|z|=1$, there are three cases:
\begin{itemize}
\item The series diverges if $p\geq 1$. As a constraint on $\a$ it yields $\a\geq 1$. This case is excluded by construction, although the limit case $\a=1$ is well-defined.
\item The series converges absolutely if $p<0$, corresponding to $\a<1-2/D$.
\item The series converges except at $z=1$ if $0\leq p<1$, giving $1-2/D\leq \a<1$.
\item When $\a=2/D$, the series diverges if $D\leq 2$, converges absolutely if $D>4$, and converges except at $z=1$ if $2<D\leq 4$. These bounds are basically unchanged if one considers the analytic continuation of $F$ to the massless case.
\end{itemize}
The standard physical setting (convergence of the propagator in the unit disk and on its boundary except at the singularity on the real $k^0$ axis) is naturally recovered only for
\be
2\leq D\leq 4,\label{bou}
\ee
where we included both extrema of the interval by analytic continuation.

Before concluding the section, we reconsider the issue of the superficial degree of divergence of Feynman graphs in the UV (see, e.g., \cite{Vi09a,Vi09b} for an introduction to the subject and references). Consider a one-particle-irreducible subdiagram with $L$ loops, $I\geq L$ internal propagators and $V$ vertices. The superficial degree of divergence $\delta$ is the canonical dimension of all these contributions. 

Each loop integral gives $[\rmd\vr(k)]=D\a$, while the propagator, in any dimension and for any value of $\a$, has $[\tilde G]=-2$. For the scalar field theory, interaction vertices do not carry dimensionality. Overall,
\be
\delta = L(D\a-2)-2(I-L)\leq L(D\a-2)\,.
\ee
When $\a=1$, one gets the standard result in $D$ dimensions. In the critical case $\a=2/D$, $\delta\leq 0$ and one has at most logarithmic divergences. When $\a<2/D$ the theory is superrenormalizable. In the case of gravity also vertices contribute \cite{Vi09a,Vi09b}, each with a factor of $2$ (number of derivatives). Then, $\delta$ is bounded by the dimension of operators which already appear in the bare action.


\section{Gravity}\label{gravy}

Having studied the properties of a scalar field on an effective fractal spacetime, we turn to gravity. Our conventions for the Levi-Civita connection, Riemann and Ricci tensors, and Ricci scalar are
\ba
&& \G^\a_{\mu\nu}\equiv \tfrac12g^{\a\b}\left[\p_{\mu} g_{\nu\b}+\p_{\nu} g_{\mu\b}-\p_\b g_{\mu\nu}\right]\,,\label{leci}\\
&& R^\a_{~\mu\b\nu}\equiv \p_\b \G^\a_{\mu\nu}-\p_\nu \G^\a_{\mu\b}+\G^\s_{\mu\nu}\G^\a_{\b\s}-\G^\s_{\mu\b}\G^\a_{\nu\s}\,,\\
&& R_{\mu\nu}\equiv R^\a_{~\mu\a\nu}\,,\qquad R\equiv R_{\mu\nu}g^{\mu\nu}\,.
\ea
The \emph{Ansatz} for the gravitational action is
\be\label{Sg}
S_g=\frac{1}{2\k^2}\int\rmd\vr(x)\,\sqrt{-g}\,\left(R-2\la-\om\p_\mu v\p^\mu v\right)\,, 
\ee
where $g$ is the determinant of the dimensionless metric $g_{\mu\nu}$, $\k^2=8\pi G$ is Newton's constant, $\la$ is a bare cosmological constant, and the term proportional to $\om$ has been added because $v$, like the other geometric field $g_{\mu\nu}$, is now dynamical. The couplings have dimension
\be
[\k^2]= 2-D\a\,,\qquad [\la]=2\,,\qquad [\om]=2D(\a-1)+4\,.
\ee
In spacetime with $D=2$ topological dimensions and trivial measure weight $v=1$, the Einstein--Hilbert action is a topological invariant and there are no dynamical degrees of freedom. This is not the case of eq.~\Eq{Sg}.

To describe the flow from the UV to the IR fixed point, we should add relevant operators also into the gravitational action. (The relevant operators in the matter sector are minimally coupled with gravity and they would not be enough.) This is done in the same way as for the matter sector, eqs.~\Eq{reno} and \Eq{mea}, with possibly different $M_g\neq M_\phi$. The effective Newton constant then runs from a UV bare dimensionless constant to an IR value
\be\label{hie}
\k^2_{\rm IR}\sim  \k^2_{\rm UV} M_g^{2-D} \,. 
\ee
Note that $\k^2_{\rm IR}$ is not necessarily the observed Newton constant $\k^2_{\rm obs}$, in which case $M_g\sim m_{\rm Pl}$ is the Planck mass. As one can see from the equations of motion, $\k^2_{\rm obs}$ will depend on the background as well as on the scale of the problem.

Those in eq.~\Eq{reno} are not all possible relevant operators. Higher-order Riemann invariants ${\rm Riem}^n$ ($[{\rm Riem}^n]=2n$, $n>1$) are irrelevant, but one could introduce lowest-order terms of the form $(1+{\rm Riem})^n$ with $n<1$. These terms might be expected in the context of fractal models, where fractional derivatives can (but not necessarily) find their natural setting.\footnote{If they were regarded as necessary, all $\p^\b$ terms should make their appearance in any sector. This is a different \emph{definition} of the theory which we will not consider.} Considering all these operators, one ends up with a term
\be
\int_0^1 \rmd n\,c(n) \left(1+\frac{{\rm Riem}}{{\rm mass}^2}\right)^n\,,
\ee
where $c(n)$ are arbitrary dimensionless coefficients. In the unrealistic case where they are all equal to 1 and ${\rm Riem}=R$, the integral can be summed explicitly to a nonpolynomial functional which admits Minkowski as a vacuum and yields the Einstein--Hilbert Lagrangian with cosmological constant at small $R$ (in mass units):
\be
\frac{R}{\ln(1+R)}= 1+\frac{R}2-\frac{R^2}{12}+O(R^3)\,.\nonumber
\ee 
This particular $f(R)$ model might be of some cosmological interest. However, it is a toy model and we shall not continue its discussion. In fact, we are interested in the equations of motion near the UV fixed point, so we will ignore relevant operators from now on ($M_g=M_\phi=0$).


\subsection{Einstein equations}\label{einseqs}

Assuming that matter is minimally coupled with gravity, the total action is
\be
S=S_g+S_{\rm m}\,,
\ee
where $S_g$ is eq.~\Eq{Sg} and $S_{\rm m}=\int \rmd\vr \sqrt{-g} \cL_{\rm m}$ is the matter action. The derivation of the Einstein equations is almost as in scalar-tensor models. We shall repeat it here to make the presentation self-contained. To find the equations of motion we need the variations
\ba
\delta \sqrt{-g}&=& -\tfrac12\,g_{\mu\nu}\,\sqrt{-g}\,\delta g^{\mu\nu}\,,\\
\delta R        &=& (R_{\mu\nu}+g_{\mu\nu}\,\B-\N_\mu\N_\nu)\,\delta g^{\mu\nu}\,,
\ea
where $\N_\nu V_\mu \equiv \p_\nu V_\mu-\G^\s_{\mu\nu}V_\s$ is the covariant derivative of a vector $V_\mu$ and the curved d'Alembertian on a scalar $\phi$ is 
\be\label{dal}
\Box\phi =\frac{1}{\sqrt{-g}}\p^\mu (\sqrt{-g}\p_\mu\phi)\,.
\ee
The Einstein equations $\delta S/\delta g^{\mu\nu}=0$ read
\ba
\Sigma_{\mu\nu}&=&\k^2T_{\mu\nu},\label{eineq}\\
\Sigma_{\mu\nu}&\equiv& R_{\mu\nu}-\frac12g_{\mu\nu}(R-2\la)+g_{\mu\nu}\frac{\B v}{v} -\frac{\N_\mu\N_\nu v}{v}\nonumber\\
&&+\om\left(\frac12 g_{\mu\nu}\p_\s v\p^\s v-\p_\mu v\p_\nu v\right)\,,\\
T_{\mu\nu} &\equiv&-\frac{2}{\sqrt{-g}}\frac{\delta S_{\rm m}}{\delta g^{\mu\nu}}=-2\frac{\p\cL_{\rm m}}{\p g^{\mu\nu}}+g_{\mu\nu}\cL_{\rm m}\,.
\ea
Taking the trace of eq.~\Eq{eineq} gives
\be\label{tra}
-\left(\frac{D}2-1\right)R+D\la+(D-1)\frac{\Box v}{v}+\left(\frac{D}2-1\right)\om\p_\mu v\p^\mu v=\k^2T_\mu^{\ \mu}.
\ee
When taking into account the variation of the total action with respect to the scalar $v$,
\be\label{veom}
R-2\la=-2\k^2\cL_{\rm m}-\om \left(2v\B v+\p_\mu v\p^\mu v\right)\,,
\ee
eq.~\Eq{tra} becomes
\ba
R+(D-1)\frac{\Box v}{v}+\om\left[D v\B v+(D-1)\p_\mu v\p^\mu v\right]&=&\k^2\left(T_\mu^{\ \mu}-D\cL_{\rm m}\right)\nonumber\\
&=&-2\k^2\,{\rm Tr}\frac{\p\cL_{\rm m}}{\p g^{\mu\nu}}.\label{tra2}
\ea

$T_{\mu\nu}$ is the stress-energy tensor of matter. Its definition determines the continuity equation \cite{wei72}. In fact, let
\be\label{infa}
\delta S_{\rm m}=\frac12\int\rmd^D x\, v\sqrt{-g}\,T^{\mu\s}\delta g_{\mu\s}+\int\rmd^D x \sqrt{-g}\,\cL_{\rm m}\delta v
\ee
be the infinitesimal variation of the matter action with respect to the external fields $\delta g_{\mu\s}$ and $\delta v$. For the infinitesimal coordinate transformation \Eq{tran}, one has
\ba
&&\delta g_{\mu\s}=g_{\nu\s}\p_\mu a^\nu+g_{\mu\nu}\p_\s a^\nu+a^\nu\p_\nu g_{\mu\s}\,,\label{va1}\\
&&\delta v=a^\nu\p_\nu v\,,\label{va2}
\ea
where we used the definition of the Lie derivative for rank-2 and rank-0 tensors. Plugging eqs.~\Eq{va1} and \Eq{va2} into \Eq{infa} and integrating by parts, we get
\be\label{infa2}
\delta S_{\rm m}=-\int\rmd^D x \,a^\nu\left[\p_\mu(v\sqrt{-g}T^{\mu}_{\ \nu})-\frac12 v\sqrt{-g}T^{\mu\s}\p_\nu g_{\mu\s}-\p_\nu v\,\sqrt{-g}\cL_{\rm m}\right]\,.
\ee
$\delta S_{\rm m}$ must vanish on shell (i.e., when the dynamical equations are satisfied).
Using the properties of the Levi-Civita connection \Eq{leci} and the definition of the covariant derivative of a rank-2 tensor,
\ba
\N_\mu T^\mu_{\ \nu}&=&\p_\mu T^\mu_{\ \nu}+\G^\mu_{\mu\s}T^\s_{\ \nu}-\G^\s_{\mu\nu}T^\mu_{\ \s}\nonumber\\
&=& \frac{1}{\sqrt{-g}}\p_\mu\left(\sqrt{-g}T^\mu_{\ \nu}\right)-\frac12 (\p_\nu g_{\mu\s})T^{\mu\s}\,,\nonumber
\ea
one finally obtains the continuity equation
\be\label{conteq}
\N_\mu (v T^{\mu}_{\ \nu})-\p_\nu v\,\cL_{\rm m}=0\,,
\ee
which generalizes eq.~\Eq{noe1}.

If matter is a scalar field, it is straightforward to see that its equation of motion $\delta S_{\rm m}/\delta\phi =0$ is eq.~\Eq{eom} with $\B$ given by eq.~\Eq{dal}, in agreement with eq.~\Eq{conteq}.

The continuity and Einstein equations are not independent because of the contracted Bianchi identities $2\N^\mu R_{\mu\nu}=\N_\nu R$ and eq.~\Eq{veom}. The divergence of ($v$ times) eq.~\Eq{eineq} correctly reproduces eq.~\Eq{conteq}. The check takes into account that in the absence of torsion the covariant derivative commutes on a scalar, $[\N_\mu,\N_\nu]v=0$, while on a vector $[\N_\mu,\N_\nu]V_\s=R_{\mu\nu\s}^{\ \ \ \ \tau}V_\tau$.


\subsection{Cosmology}\label{cosmo}

With the notable difference that matter is nonminimally coupled with the scalar $v$, the equations of motion are similar to those of Brans--Dicke theory \cite{BD}, which is well constrained by large-scale observations \cite{DMT}. This fact and the foreign physical setting lead to an altogether different dynamics.

In this section we specialize to a Friedmann--Robertson--Walker (FRW) line element
\be
\rmd s^2=g_{\mu\nu}\rmd x^\mu\rmd x^\nu=-\rmd t^2+a(t)^2\tilde g_{ij}\rmd x^i\rmd x^j\,,
\ee
where $t$ is synchronous time, $a(t)$ is the scale factor and
\be
\tilde g_{ij}\rmd x^i \rmd x^j=  \frac{\rmd r^2}{1-\textsc{k}\,r^2}+r^2\rmd\Om^2_{D-2}
\ee
is the line element of the maximally symmetric $(D-1)$-dimensional space $\tilde\Sigma$ of constant sectional curvature $\textsc{k}$ (equal to $-1$ for an open universe, 0 for a flat universe and $+1$ for a closed universe with radius $a$). In four dimensions, $\rmd\Om^2_{2}=\rmd\theta^2+\sin^2\theta\rmd\varphi^2$. Quantities built up with the spatial metric $\tilde g_{ij}$ will be decorated with a tilde. On this background, the only nonvanishing Levi-Civita and Ricci components are
\be
\G_{ij}^0=Hg_{ij},\qquad \G_{i 0}^j=H\delta_i^j,\qquad \G_{ij}^k=\tilde\G_{ij}^k\,,
\ee
and
\ba
&& R_{00}   = -(D-1)(H^2+\dot H)\,,\\
&& R_{ij} =\tilde R_{ij}+[(D-1)H^2+\dot H]g_{ij},\qquad
\tilde R_{ij}=\frac{2\textsc{k}}{a^2}g_{ij}\,,\\
&& R = (D-1)\left(\frac{2\textsc{k}}{a^2}+DH^2+2\dot{H}\right),
\ea
where 
\be
H\equiv \frac{\dot{a}}{a}
\ee
is the Hubble parameter (not to be confused with the Hamiltonian $H$ of section \ref{eoh}) and we have exploited the symmetries of $\tilde\Sigma$ \cite{wei72}.

For simplicity we consider a perfect fluid (zero heat flow and anisotropic stress) as the only content of the universe:
\be
T_{\mu\nu}=(\rho+p)\,u_\mu u_\nu+p\,g_{\mu\nu}\,,
\ee
where $\rho=T_{00}$ and $p=T_i^{\ i}/(D-1)$ are the energy density and pressure of the fluid and $u^\mu=(1,0,\dots,0)^\mu$ is the unit timelike vector ($u_\mu u^\mu=-1$) tangent to a fluid element's worldline. We also take a timelike fractal $v=v(t)$.

The $00$ component of the Einstein equations \Eq{eineq} is
\be\label{00}
\left(\frac{D}2-1\right)H^2+H\frac{\dot v}{v}-\frac{1}{2}\frac{\om}{D-1}\dot v^2=\frac{\k^2}{D-1}\rho +\frac{\la}{D-1}-\frac{\textsc{k}}{a^2}\,,
\ee
while combining that with the trace equation \Eq{tra} one obtains
\be\label{fr2}
\frac{\Box v}{v}-(D-2)\left(H^2+\dot H-H\frac{\dot v}{v}+\frac{\om}{D-1}\dot v^2\right)+\frac{2\la}{D-1}=\frac{\k^2}{D-1}\left[(D-3)\rho+(D-1)p\right].
\ee
Other useful expressions can be found by suitable combinations of the dynamical equations. From eq.~\Eq{tra2},
\be\label{trn}
R+(D-1)\frac{\Box v}{v}+\om\left[D v\B v-(D-1)\dot v^2\right]=-\k^2(\rho+p)\,,
\ee
and eq.~\Eq{veom} one gets
\be\label{tra3}
2\la+(D-1)\frac{\Box v}{v}+(D-2)\om (v\B v-\dot v^2)=-\k^2(\rho-p)\,,
\ee
while from the $00$ component of \Eq{eineq} and eq.~\Eq{veom},
\be\label{trm}
H^2+\dot H-H\frac{\dot v}{v}+\frac{\om}{D-1}v\B v=-\frac{\k^2}{D-1}(\rho+p)\,.
\ee
If $\rho+p\neq 0$, one can combine eqs.~\Eq{trn} and \Eq{trm} to get a purely gravitational constraint:
\be\label{grac}
\dot{H}+(D-1)H^2+\frac{2\textsc{k}}{a^2}+\frac{\Box v}{v}+H\frac{\dot v}{v}+\om (v\B v-\dot v^2)=0\,.
\ee
The continuity equation \Eq{conteq} contracted with $-u^\nu$ is
\be\label{pippo}
\dot\rho+\left[(D-1)H+\frac{\dot v}{v}\right](\rho+p)=0\,,
\ee
where we used the definition of proper-time derivative, $u_\mu\N^\mu=\dot{}$, and the Hubble expansion $\Theta\equiv\N^\mu u_\mu=(D-1)H$ in covariant formalism \cite{Haw66,Ell71,EB}. Note that this is not the volume expansion as in standard general relativity, as the latter is actually
\be
\tilde\Theta=\frac{\rmd \ln (a^{D-1}v)}{\rmd t}\,,
\ee
which is the square bracket in eq.~\Eq{pippo}. For a barotropic fluid $p=w\rho$, the continuity equation is solved by $\rho\sim (a^{D-1}v)^{-(1+w)}$, up to some dimensionful prefactor. In general $w$ is not a constant and we define the effective barotropic index
\be\label{bari}
w(t)\equiv \frac{p}{\rho}\,.
\ee
The scalar field is a particular case of perfect fluid, with $p=\cL_\phi=\dot\phi^2/2-V$, $\rho=\dot\phi^2/2+V$, and $u_\mu=-\p_\mu\phi/\dot\phi$ \cite{Mad88}. Equation \Eq{pippo} becomes eq.~\Eq{eom},
\be
\ddot\phi+\left[(D-1)H+\frac{\dot v}{v}\right]\dot\phi+V'=0\,.
\ee

When $v=1$ and $D=4$, we recover the standard Friedmann equations in four dimensions, eqs.~\Eq{00} and \Eq{fr2} (no gravitational constraint):
\ba
&&H^2=\frac{\k^2}{3}\rho+\frac{\la}{3}-\frac{\textsc{k}}{a^2}\,,\label{fr1}\\
&&H^2+\dot{H}=-\frac{\k^2}{6}(3p+\rho)+\frac{\la}{3}\,.
\ea
On the other hand, for the measure weight
\be\label{vt}
v=t^{-\b}\,,
\ee
where $\b$ is given by eq.~\Eq{beta}, the gravitational constraint is switched on.
Then, the above equations should be taken \emph{cum grano salis}. The UV regime, in fact, describes short scales at which inhomogeneities should play some role. If these are small, the modified Friedmann equations define a background for perturbations rather than a self-consistent dynamics.

Modulo this caveat, we can look at flat ($\textsc{k}=0$) background solutions in the deep UV regime with no cosmological constant. The gravitational constraint \Eq{grac} is a Riccati equation in $H$ which, together with the useful formul\ae\
\be
H\frac{\dot v}{v}= -H\frac{\b}{t}\,,\qquad \frac{\Box v}{v}=\frac{\b}{t}\left[(D-1)H-\frac{1+\b}{t}\right],
\ee
fixes almost completely the background expansion, regardless the matter content. A direct consequence of this overdetermination of the dynamics is that there are no vacuum solutions ($\rho=0=p$). This might not happen for other fractal profiles than eq.~\Eq{vt}, but at early times the measure must scale as eq.~\Eq{vt}. This feature, therefore, is robust.

Let us consider the cases $\om=0$ and $\om\neq 0$ separately. When $\b=D-2=2$ (UV regime), the $\om=0$ solution is
\bs\ba
a(t) &=& \frac{(t^9+c)^{1/3}}{t^2}\,,\\
H(t) &=& \frac{t^9-2c}{t (t^9+c)}\,,\\
\epsilon &\equiv& -\frac{\dot H}{H^2} =\frac{[t^9-(14+3\sqrt{22})c][t^9-(14-3\sqrt{22})c]}{(t^9-2c)^2}\,,
\ea
where $c$ is an integration constant. The energy density and pressure which solve all the equations simultaneously are
\ba
\rho &=&-\frac{3}{\k^2}\frac{(t^9+4c)(t^9-2c)}{t^2(t^9+c)^2}\,,\\
p    &=&-\frac{3}{\k^2}\frac{[t^9+(\sqrt{10}-3)\sqrt{10}c][t^9+(\sqrt{10}+3)\sqrt{10}c]}{t^2(t^9+c)^2}\,. 
\ea\es

These expressions are sufficient to characterize three cases:
\begin{itemize}
\item $c>0$: The scale factor decreases ($H<0$) from $t=0$ until $t=t_*\equiv(2c)^{1/9}$, where the universe bounces ($H_*=0$, $a_*=3^{1/3}2^{-2/9}c^{1/9}$). From $t=t_*$ to $t=t_1\equiv [(14+3\sqrt{22})c]^{1/9}$, the universe expands in superacceleration ($\epsilon<0$), while for $t>t_1$ the expansion is only accelerated. The energy density $\rho$ is negative for $t>t_*$, while the pressure $p$ is always negative.
\item $c=0$: Linear (decelerating) expansion, $a=t$, while $\rho=p<0$ always.
\item $c<0$: The universe expands in deceleration from a big bang event at $t=t_0\equiv |c|^{1/9}$. The energy density and pressure are negative for $t>|4c|^{1/9}$ and $t>|(\sqrt{10}+3)\sqrt{10}c|^{1/9}$, respectively.
\end{itemize}
All these scenarios need a matter component with non-positive definite energy density, so they are excluded if only ordinary matter is allowed. We envisage four simple modifications of this result. One is to change the flat prescription $\textsc{k}=0$. Another is to consider the above formul\ae\ only asymptotically, since the simple measure profile eq.~\Eq{vt} is certainly valid only at early times. The only case where the universe expands at small $t$ with $\rho>0$ is $c<0$, for which one has superstiff matter ($w(t)>1$). A third option is to allow for a nonzero geometric contribution $U(v)$, a potential for $v$.

A fourth possibility is that $\om\neq 0$, which does lead to interesting cosmology.
There is only one real solution to the gravitational constraint, namely,
\ba
a(t) &=& \frac{1}{t^2}\Phi\left(\frac{11}{4};\frac{13}{4};\frac{3\om}{2t^4}\right)^{1/3}\,,\\
H(t) &=& -\frac{2}{t}-\frac{22\om}{13 t^5} \frac{\Phi\left(\tfrac{15}{4};\tfrac{17}{4};\frac{3\om}{2t^4}\right)}{\Phi\left(\tfrac{11}{4};\tfrac{13}{4};\frac{3\om}{2t^4}\right)}\,,
\ea
where $\Phi$ (also denoted as $_1F_1$ or $M$) is Kummer's confluent hypergeometric function of the first kind:
\be\label{kumm}
\Phi(a;\,b;\,z)\equiv\frac{\Gamma(b)}{\Gamma(a)}\,\sum_{n=0}^{+\infty} \frac{\Gamma(a+n)}{\Gamma(b+n)}\, \frac{z^n}{n!}\,.
\ee
The expressions for $\rho$ and $p$ are
\ba
\rho &=& \frac{2(2\om+3t^4)(3\om+4t^4)}{t^{10}}+\frac{48\om^2}{13^2t^{10}}\frac{\Phi\left(\tfrac{11}{4};\tfrac{17}{4};\frac{3\om}{2t^4}\right)^2}{\Phi\left(\tfrac{11}{4};\tfrac{13}{4};\frac{3\om}{2t^4}\right)^2}\nonumber\\
&&-\frac{24\om(2\om+3t^4)}{13t^{10}}\frac{\Phi\left(\tfrac{11}{4};\tfrac{17}{4};\frac{3\om}{2t^4}\right)}{\Phi\left(\tfrac{11}{4};\tfrac{13}{4};\frac{3\om}{2t^4}\right)}\,,\\
p &=& -\frac{2(\om+3t^4)(6\om+5t^4)}{t^{10}}+\frac{48\om^2}{13^2t^{10}}\frac{\Phi\left(\tfrac{11}{4};\tfrac{17}{4};\frac{3\om}{2t^4}\right)^2}{\Phi\left(\tfrac{11}{4};\tfrac{13}{4};\frac{3\om}{2t^4}\right)^2}\,.
\ea
We can use the asymptotic forms of $\Phi$ to have some semi-analytic insight of the system. When $z\to -\infty$,
\be
\Phi(a;\,b;\,z) \stackrel{z\to-\infty}{\sim}
\frac{\Gamma(b)}{\Gamma(b-a)}(-z)^{-a}+(-z)^{-a} \sum_{n=1}^{N}\frac{\Gamma(a+n)}{\Gamma(a)}\frac{\Gamma(b)}{\Gamma(b-a-n)}\frac{z^{-n}}{n!}\,,\label{dive}
\ee
where $N$ is some finite order. On the other hand,
\be
\Phi(a;\,b;\,z) \stackrel{z\to+\infty}{\sim} \frac{\Gamma(b)}{\Gamma(a)}\,\rme^{z}z^{a-b}+\rme^{z}z^{a-b} \sum_{n=1}^{N}\frac{\Gamma(b-a+n)}{\Gamma(b-a)}\frac{\Gamma(b)}{\Gamma(a-n)}\frac{(-z)^{-n}}{n!}\,.\label{dive2}
\ee
At late times the scale factor decreases and the fluid behaves effectively as phantom matter ($w<-1$):
\bs\ba
a &\stackrel{t\to +\infty}{\sim}& \frac{1}{t^2}\,,\qquad H \sim -\frac{2}{t}\,,\\
\epsilon &\sim& -\frac{1}{2}\,,\\
\rho &\sim& \frac{24}{t^2}\,,\qquad p \sim -\frac{30}{t^2}\,,\\
w &\sim& -\frac{5}{4}\,.
\ea\es
At early times we must distinguish between positive and negative $\om$. For $\om>0$, the universe is contracting and the fluid behaves like an effective cosmological constant:
\bs\ba
a &\stackrel{t\to 0}{\sim}& {\rm const}\times\frac{\rme^{\frac{\om}{2t^4}}}{t^{4/3}}\,,\qquad
H \sim -\frac{2\om}{t^5}\,,\\
\epsilon &\sim& -\frac{5t^4}{2\om}\,,\\
\rho &\sim& \frac{12\om^2}{t^{10}}\,,\qquad p \sim -\frac{12\om^2}{t^{10}}\,,\\
w &\sim& -1\,.
\ea\es
These quantities are plotted in figures \ref{fig1} and \ref{fig2} for $\om=+1$.
\FIGURE{
\includegraphics[width=7.4cm]{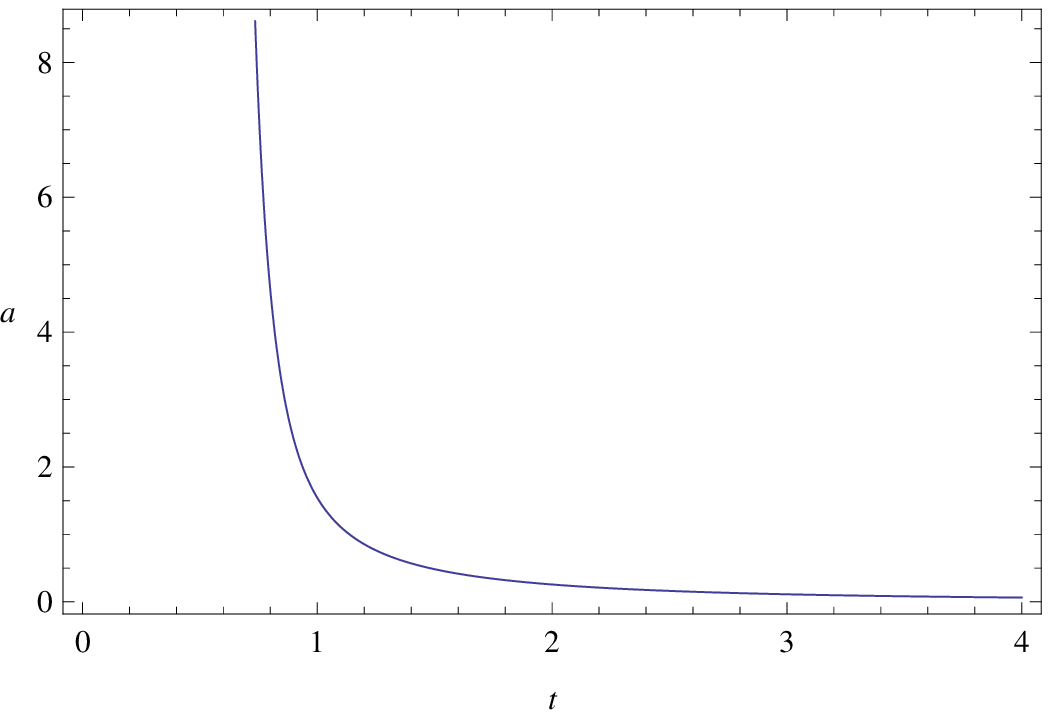}
\includegraphics[width=7.4cm]{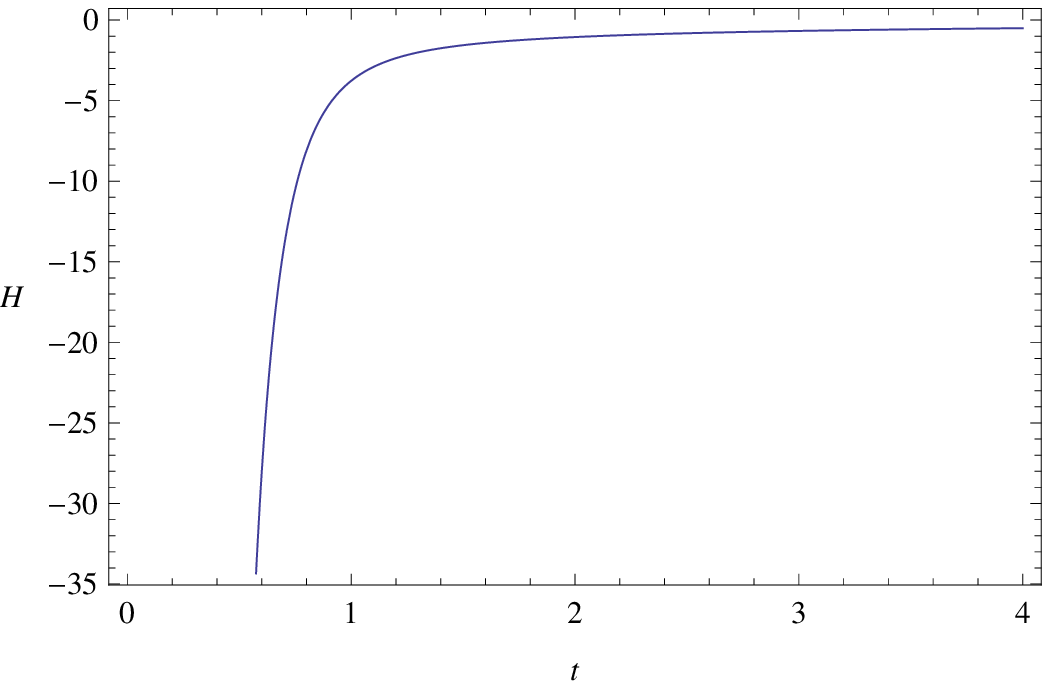}
\includegraphics[width=7.4cm]{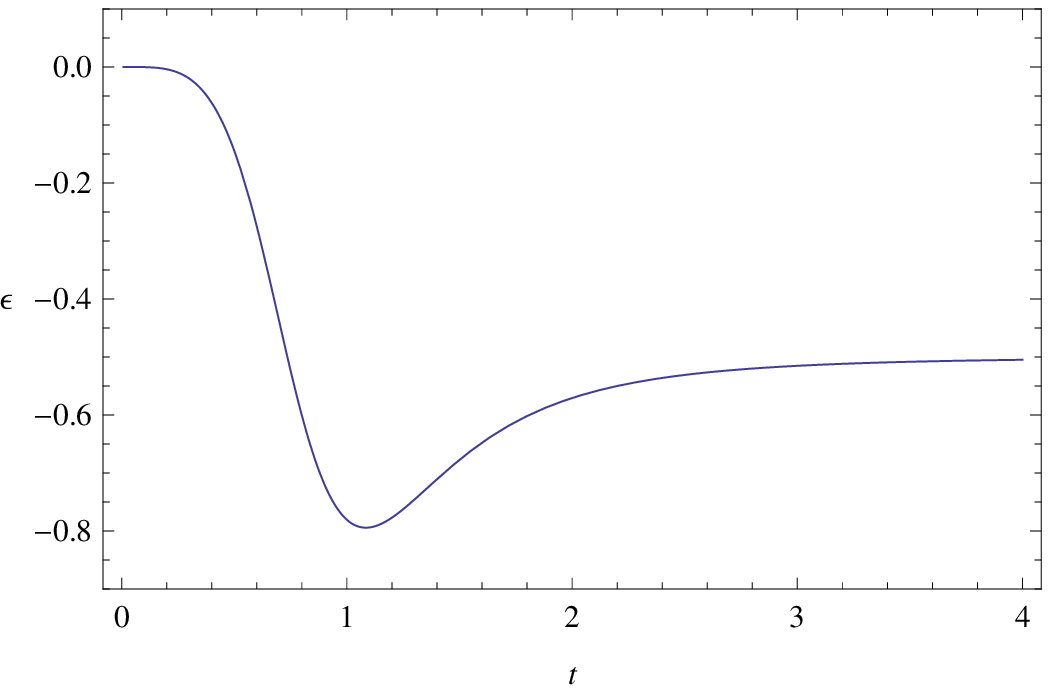}
\includegraphics[width=7.4cm]{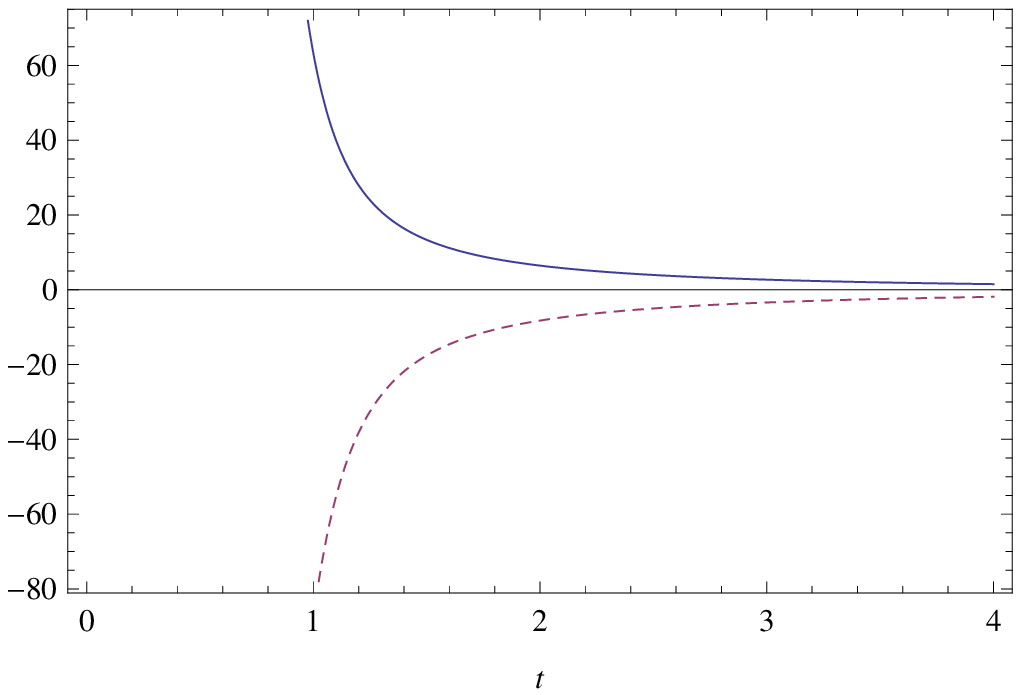}
\caption{\label{fig1} The scale factor $a$, Hubble parameter $H$, slow-roll parameter $\epsilon$, energy density $\rho$ (thick line) and pressure $p$ (dashed line) for $\om=+1$.}}
\FIGURE{
\includegraphics[width=9cm]{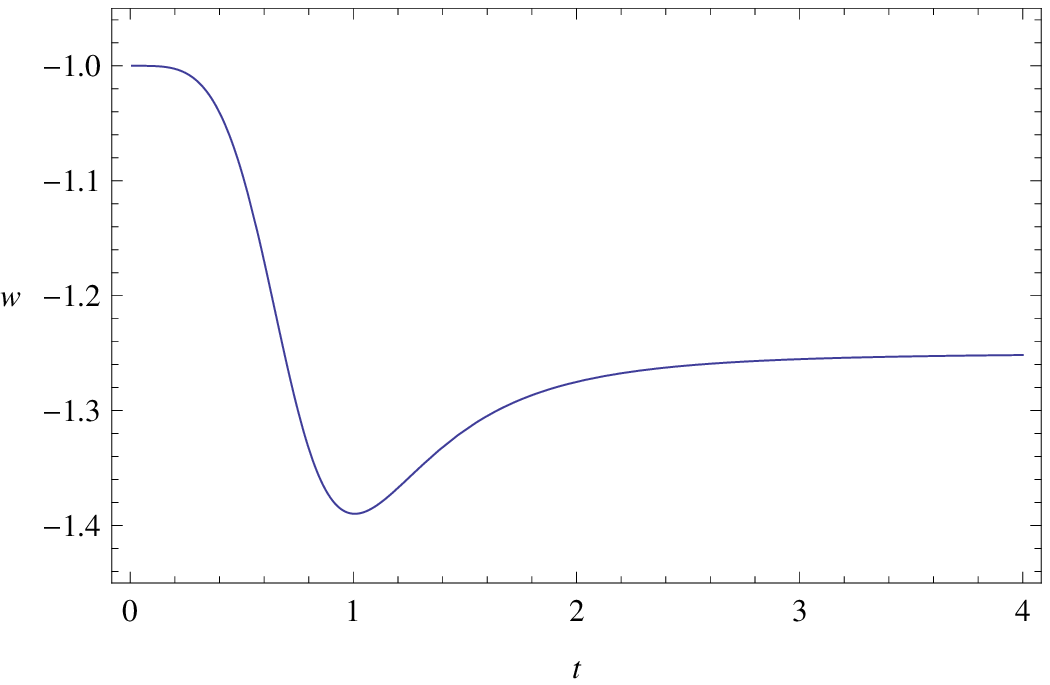}
\caption{\label{fig2} The equation of state $w=p/\rho$ for $\om=+1$.}}

For $\om<0$, at early times the universe expands and accelerates, even if the perfect fluid is stiff:
\bs\ba
a &\stackrel{t\to 0}{\sim}& {\rm const}\times t^{5/3}\,,\qquad H \sim \frac{5}{3t}\,,\\
\epsilon &\sim& \frac{3}{5}\,,\\
\rho &\sim& \frac{2|\om|}{t^6}\,,\qquad p \sim \frac{2|\om|}{t^6}\,,\\
w &\sim& 1\,.
\ea\es
These are plotted in figures \ref{fig3} and \ref{fig4} for $\om=-1$.
\FIGURE{
\includegraphics[width=7.4cm]{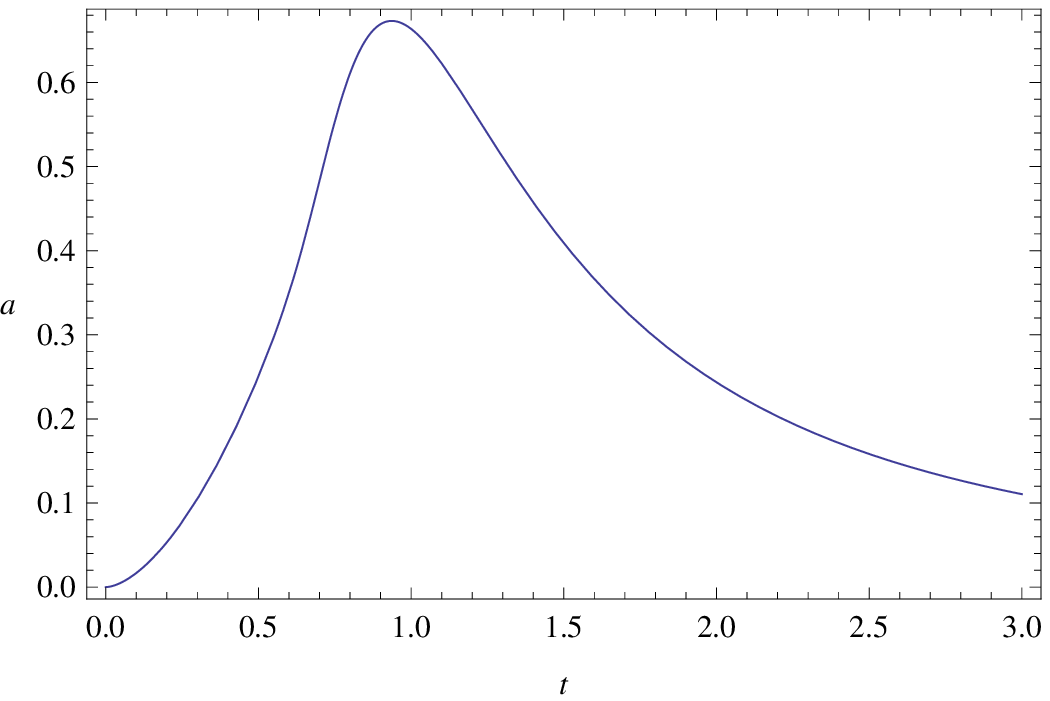}
\includegraphics[width=7.4cm]{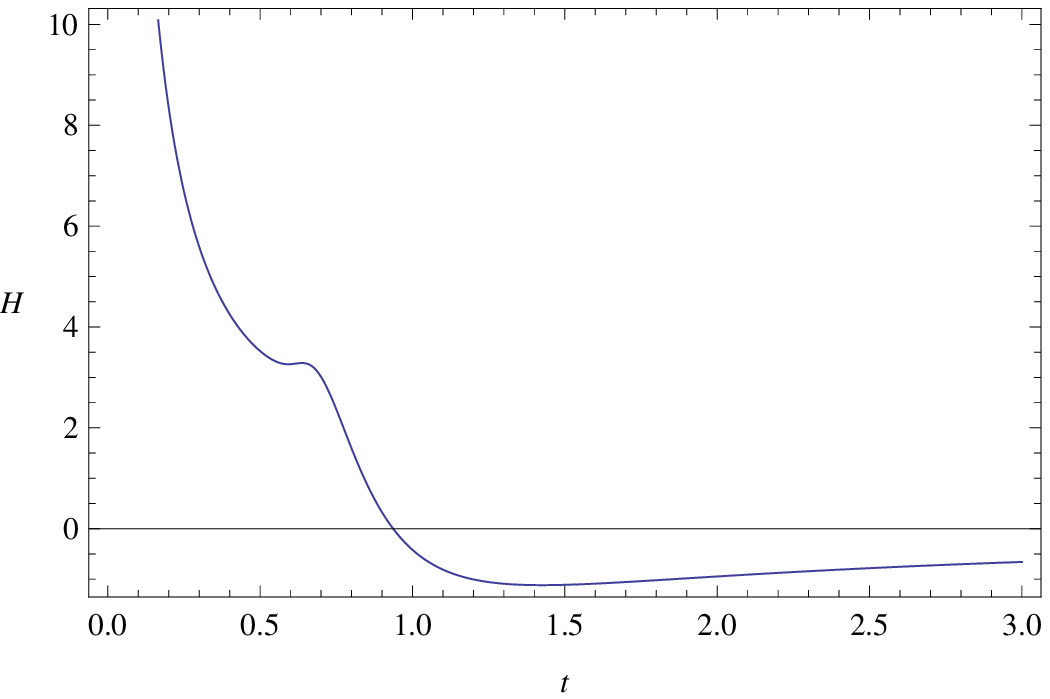}
\includegraphics[width=7.4cm]{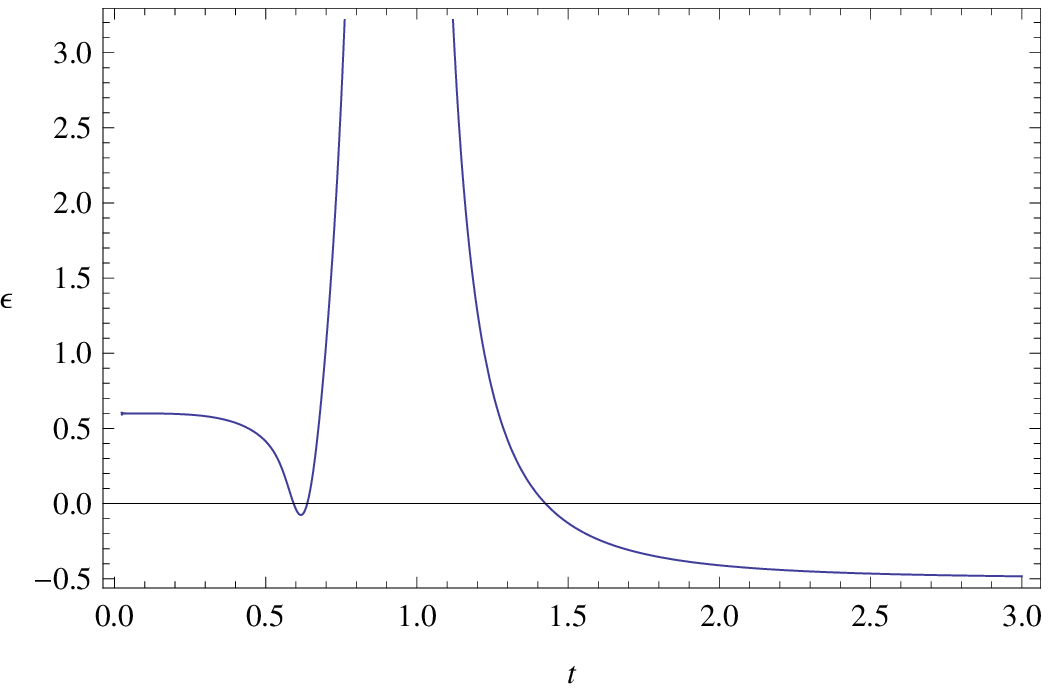}
\includegraphics[width=7.4cm]{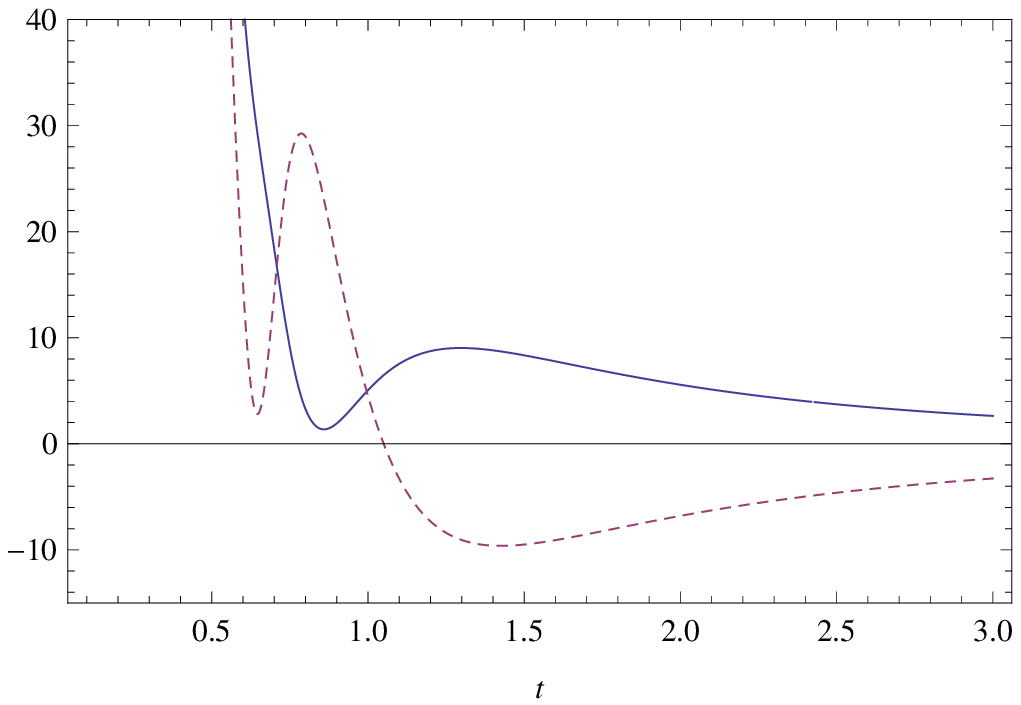}
\caption{\label{fig3} The scale factor $a$, Hubble parameter $H$, slow-roll parameter $\epsilon$, energy density $\rho$ (solid line) and pressure $p$ (dashed line) for $\om=-1$.}}
\FIGURE{
\includegraphics[width=9cm]{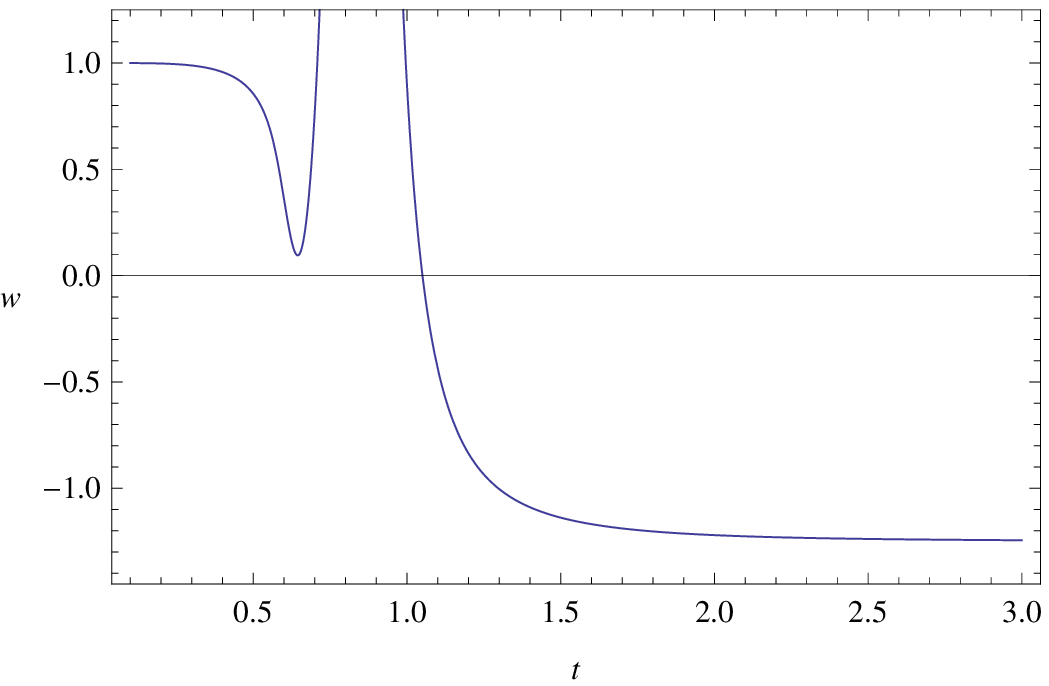}
\caption{\label{fig4} The equation of state $w=p/\rho$ for $\om=-1$. The maximum, which lies beyond the frame, is $w\approx 17.383$ at $t\approx 0.851$.}}

The null energy condition is violated, so none of these scenarios can be realized by an ordinary scalar field, for which $\rho+p=\dot\phi^2\geq 0$. The system is invariant under time reversal, so the model with $\om>0$, run backwards in time, also describes an expanding superaccelerating ($\epsilon<0$) universe filled with phantom matter. Contrary to standard general relativity, the energy density \emph{increases}. This can be explained by recalling that the fractal geometry causes dissipation. The observer, in this case, experiences an incoming flux of energy from the four-dimensional bulk.

The model with $\om<0$ is a universe which expands in acceleration. At some point the expansion is quasi de Sitter (notice the small plateau of $H$ in figure \ref{fig3}), and after a brief period of superacceleration the cosmic expansion decelerates, stops, and reverts. Again, the fluid behaves unusually: during the accelerated expansion the equation of state is between dust and stiff matter ($0<w(t)<1$), while the energy density decreases. Then, after a non-monotonic transitory period around the inversion point, the equation of state violates first the strong and then the null energy condition. Eventually the energy density still decreases even during the cosmic contraction, because the world-fractal dissipates into the bulk.

The overall cosmological picture may be regarded as problematic for any value of $\om$. On one hand, there are no vacuum solutions. On the other hand, solutions with matter require fluids violating most or all energy conditions, thus signalling an unstable or unrealistic field (however, see below). The $\om<0$ case does so explicitly, since $v$ is a ghost.

This might be cured by allowing a nonzero intrinsic curvature, more complicated matter profiles, or a nontrivial potential for $v$.\footnote{Different geometric profiles would not work. For $\om>0$, the null energy condition is violated at early times ($|t|> 1$ in the backwards model) where eq.~\Eq{vt} holds. For $\om<0$, $v$ is a ghost by definition.} However, the most natural possibility is that a classical FRW background, either exact or linearly perturbed, is not realistic. Then, one would have to treat the UV limit as highly inhomogeneous. This is not at all unexpected, as we are dealing with quantum scales where the minisuperspace equations (maximal symmetry) are likely to fail.


\section{Conclusions and future developments} \label{disc}

There are several avenues of investigation left to explore. Here we mention just three. 
\begin{itemize}
\item\emph{Quantum field theory.} Many aspects of the field theory have yet to be fully understood: among the most important are renormalization, the hierarchy problem and the physical significance of the UV propagator and the ``natural'' bound eq.~\Eq{bou}. Even in the Minkowski embedding, the fractal structure may have interesting properties we have not discussed here. For instance, because of violation of translation invariance the Fermi frame does not exist and, as in general relativity, parity is strictly a local symmetry.

Other formulations of the theory including fractional derivatives would modify the propagator and dissipation properties and, pending a suitable definition of the kinetic terms, they could lead to a more transparent physics. Also, for simplicity we have defined an action on an embedding spacetime without boundaries, but a very interesting alternative is to consider scenarios with boundaries; for example, a field theory defined on $\mathbb{R}^D_+$ would be closer in spirit to the unilateral fractional and Weyl integrals of fractal classical mechanics, eq.~\Eq{rli3} with $\bar t\to+\infty$. The propagator and several other features should change accordingly but, \emph{mutatis mutandis}, the main idea of a fractal universe would still be valid and maintain the same motivations.

\item\emph{Cosmology.} If the system quickly flows to the IR fixed point, dissipative effects might be negligible on cosmological spacetime scales, with a notable exception. At late times, an imprint of the nontrivial short-scale geometry might survive as a cosmological constant of purely gravitational origin (pressureless matter does not contribute to it). 
A detailed study may reveal whether the behaviour of the effective cosmological constant \Eq{Lam} is compatible with observations.

On the other hand, the UV regime may be relevant in the early universe, especially during inflation. We have seen that the flat background dynamics cannot be realized by a scalar field. It would be interesting to study the consequences of a nonvanishing intrinsic curvature or potential $U(v)$. If, even in that case, ordinary matter (in particular, a scalar field) were not allowed, then one would have to abandon maximal symmetry and standard perturbative techniques of inflationary cosmology.\footnote{That symmetry reduction may not be justified is also suggested by the fact that a purely homogeneous power-law fractal measure $v\sim t^{-\b}$ blows up at a non-integrable singularity, if $\b>1$: $\int_\epsilon^T \rmd t t^{-\b}\propto T^{1-\b}-\epsilon^{1-\b}$, diverging in the limit $\epsilon\to 0$. One can of course consider toy models where $\b<1$, which however do not address the renormalizability problem.} However, an appealing alternative is to assume that matter is actually a condensate field stemming from a fermionic sector. It is known that a condensate violates the null energy condition, its mass-gap effective energy density being negative in certain regimes \cite{AB1,ABC}. The physics of condensation is far from being exotic and is under good control. It would be interesting to see whether a Dirac sector with four-fermion interaction is renormalizable on a fractal and undergoes a condensation phase.

\item\emph{Extra dimensions.} In applications of the model we assumed that the topological dimension of embedding spacetime is $D=4$. An interesting alternative is a universe with extra topological dimensions, $D>4$. In this scenario the value of $\a$ would change and the IR limit should be realized in combination with a suitable compactification mechanism.
\end{itemize}


\begin{acknowledgments}
The author thanks R.~Gurau for discussions and is grateful to G.~Nardelli for invaluable comments and discussions.

\noindent {\bf Open Access.} This article is distributed under the terms of the Creative Commons
Attribution Noncommercial License which permits any noncommercial use, distribution,
and reproduction in any medium, provided the original author(s) and source are credited.
\end{acknowledgments}


\end{document}